\pdfoutput = 1

\documentclass[%
reprint,
superscriptaddress,
amsmath,
amssymb,
aps,
prb,
floatfix,
]{revtex4-2}

\usepackage{soul}
\usepackage{graphicx}% Include figure files
\usepackage{dcolumn}% Align table columns on decimal point
\usepackage{bm}% bold math
\usepackage{amsthm}
\usepackage{enumitem}
\usepackage{array}
\usepackage{tabularx}
\usepackage{siunitx}
\usepackage{placeins}
\usepackage{float}
\usepackage{hyperref}
\usepackage{amsmath}
\usepackage{amssymb}
\usepackage{scalerel}
\usepackage[subrefformat=parens,labelformat=parens, caption=false, config, labelfont=normalsize]{subfig}
\usepackage[capitalise]{cleveref}
\usepackage{comment}
\usepackage{multirow}
\usepackage{physics}
\usepackage{makecell}
\usepackage[commandnameprefix=always,nocomment]{changes}
\Crefname{figure}{Fig.}{Figs.}

\DeclareSIUnit\torr{Torr}
\let\lctau\tau % save the lowercase of '\tau'
\renewcommand{\tau}{\scalerel*{\lctau}{X}}
\newcommand{\A}[1]{A_{\mathrm{ex,}#1}}

\newcommand{\M}[1]{\vec{M}_{#1}}

\newcommand{\spin}[1]{\vec{S}_\mathrm{#1}}
\newcommand{\Ms}[1]{M_{\mathrm{s,}#1}}
\newcommand{\Hto}{\vec{H}_{2\rightarrow1}}
\newcommand{\Tto}{\vec{\tau}_{2\rightarrow1}}
\newcommand{\uveci}{{\bm{\hat{\textnormal{\bfseries\i}}}}}
\newcommand{\uvecj}{{\bm{\hat{\textnormal{\bfseries\j}}}}}
\newcommand{\uveck}{{\bm{\hat{\textnormal{\bfseries k}}}}}
\newcommand{\coeighty}{\text{Co}_{85}\text{Ru}_{15}}
\newcommand{\coninety}{\text{Co}_{90}\text{Ru}_{10}}

\begin{document}

\preprint{APS/123-QED}

%%%%%%%%%%%%%%%%%%%%%%%%%%%%%%%%%%%%%%%%%%%%%%%%%%%%%%%%%%%%%%
% FRONT MATTER 
%%%%%%%%%%%%%%%%%%%%%%%%%%%%%%%%%%%%%%%%%%%%%%%%%%%%%%%%%%%%%%

% Manuscript Title:\\
\title{Modeling magnetization reversal in multilayers with interlayer exchange coupling}

\author{Elliot Wadge}
\email{elliot.wadge@mail.mcgill.ca}
\affiliation{Department of Physics, Simon Fraser University, Burnaby, British Columbia V5A 1S6, Canada}

\author{Afan Terko}
\email{ata159@sfu.ca}
\affiliation{Department of Physics, Simon Fraser University, Burnaby, British Columbia V5A 1S6, Canada}

\author{George Lertzman-Lepofsky}
\email{gmlertzm@sfu.ca}
\affiliation{Department of Physics, Simon Fraser University, Burnaby, British Columbia V5A 1S6, Canada}

\author{Paul Omelchenko}
\affiliation{Department of Physics, Simon Fraser University, Burnaby, British Columbia V5A 1S6, Canada}

\author{Bret Heinrich}
\affiliation{Department of Physics, Simon Fraser University, Burnaby, British Columbia V5A 1S6, Canada}

\author{Manuel Rojas}
\affiliation{Department of Physics, Simon Fraser University, Burnaby, British Columbia V5A 1S6, Canada}

\author{Erol Girt}
\email{egirt@sfu.ca}
\affiliation{Department of Physics, Simon Fraser University, Burnaby, British Columbia V5A 1S6, Canada}

%%%%%%%%%%%%%%%%%%%%%%%%%%%%%%%%%%%%%%%%%%%%%%%%%%%%%%%%%%%%%%
% ABSTRACT
%%%%%%%%%%%%%%%%%%%%%%%%%%%%%%%%%%%%%%%%%%%%%%%%%%%%%%%%%%%%%%

\begin{abstract}

Spin spirals form inside the magnetic layers of antiferromagnetic and noncollinearly-coupled magnetic multilayers in the presence of an external field. This spin structure can be modeled to extract the direct exchange stiffness of the magnetic layers and the strength of the interlayer exchange coupling across the spacer layer. In this article, we discuss three models which describe the evolution of the spin spiral with the strength of the external magnetic field in these coupled structures: discrete energy, discrete torque, and continuous torque. These models are expanded to accommodate multilayers with any number of ferromagnetic layers, any combination of material parameters, and asymmetry. We compare their performance when fitting to the measured magnetization data of a range of sputtered samples with one or multiple ferromagnetic layers on either side of the spacer. We find that the discrete models produce better fits than the continuous for asymmetric and multi-ferromagnetic structures and exhibit much better computational scaling with high numbers of atomic layers than the continuous model. For symmetric, single-layered structures, the continuous model produces the same fit statistics and outperforms the discrete models. Lastly, we demonstrate methods to use interfacial layers to measure the exchange stiffness of magnetic layers with low interlayer exchange coupling. An open-access website has been provided to allow the fitting of magnetization as a function of field in arbitrary coupled structures using the discrete energy model.

\end{abstract}

\maketitle

%%%%%%%%%%%%%%%%%%%%%%%%%%%%%%%%%%%%%%%%%%%%%%%%%%%%%%%%%%%%%%
% INTRODUCTION
%%%%%%%%%%%%%%%%%%%%%%%%%%%%%%%%%%%%%%%%%%%%%%%%%%%%%%%%%%%%%%

\section{\label{sec: Introduction}Introduction}

Interlayer exchange coupling (IEC) is the interaction between the magnetizations of two ferromagnetic (FM) layers across a nonmagnetic metallic spacer layer (SL) \cite{grunberg_layered_1986, parkin_systematic_1991}. In smooth film structures, this coupling oscillates between antiferromagnetic and ferromagnetic as a function of the spacer layer's thickness. IEC is incorporated in nearly all thin-film magnetic devices to improve their magnetic properties \cite{duine_synthetic_2018}. In magnetic sensors and solid-state memory, antiferromagnetic IEC reduces the demagnetization field in the device, increasing thermal stability and improving resistance to field- and spin-polarized-current-induced switching \cite{fullerton_spintronics_2016, apalkov_magnetoresistive_2016, bhatti_spintronics_2017, arora_magnetic_2017}.

The primary interest of previous research has been in bilinear coupling, where the energy per area is linear in both directions of magnetization $\hat{m}_i$: $E/A = J_1 \hat{m}_1 \cdot \hat{m}_2$ \cite{stiles_interlayer_2005}. Recent findings, however, indicate that alloying nonmagnetic metals with ferromagnetic elements in the SL \cite{nunn_control_2020, abert_origin_2022, nunn_controlling_2023, besler_noncollinear_2023} can induce not only strong bilinear coupling, but also large \textit{biquadratic} coupling between FM layers. Unlike previous sources of biquadratic coupling, this approach enables fine control over both collinear and noncollinear alignments between the magnetic moments of adjacent ferromagnetic layers, i.e., any alignment between $0^\circ$ and $180^\circ$ \cite{lertzman-lepofsky_energy_2024}. Such alignment allows the development and fabrication of novel magnetic thin-film devices with the potential to outperform state-of-the-art technologies \cite{zhou_spin-torque_2008, sbiaa_magnetization_2013, matsumoto_spin-transfer-torque_2015}. The energy per unit area of the biquadratic coupling is quadratic in both directions of magnetization: $E/A = J_2 \left(\hat{m}_1 \cdot \hat{m}_2\right)^2$ \cite{stiles_interlayer_2005}. 

\begin{figure}[ht]
    \centering
    \includegraphics[width=\linewidth]{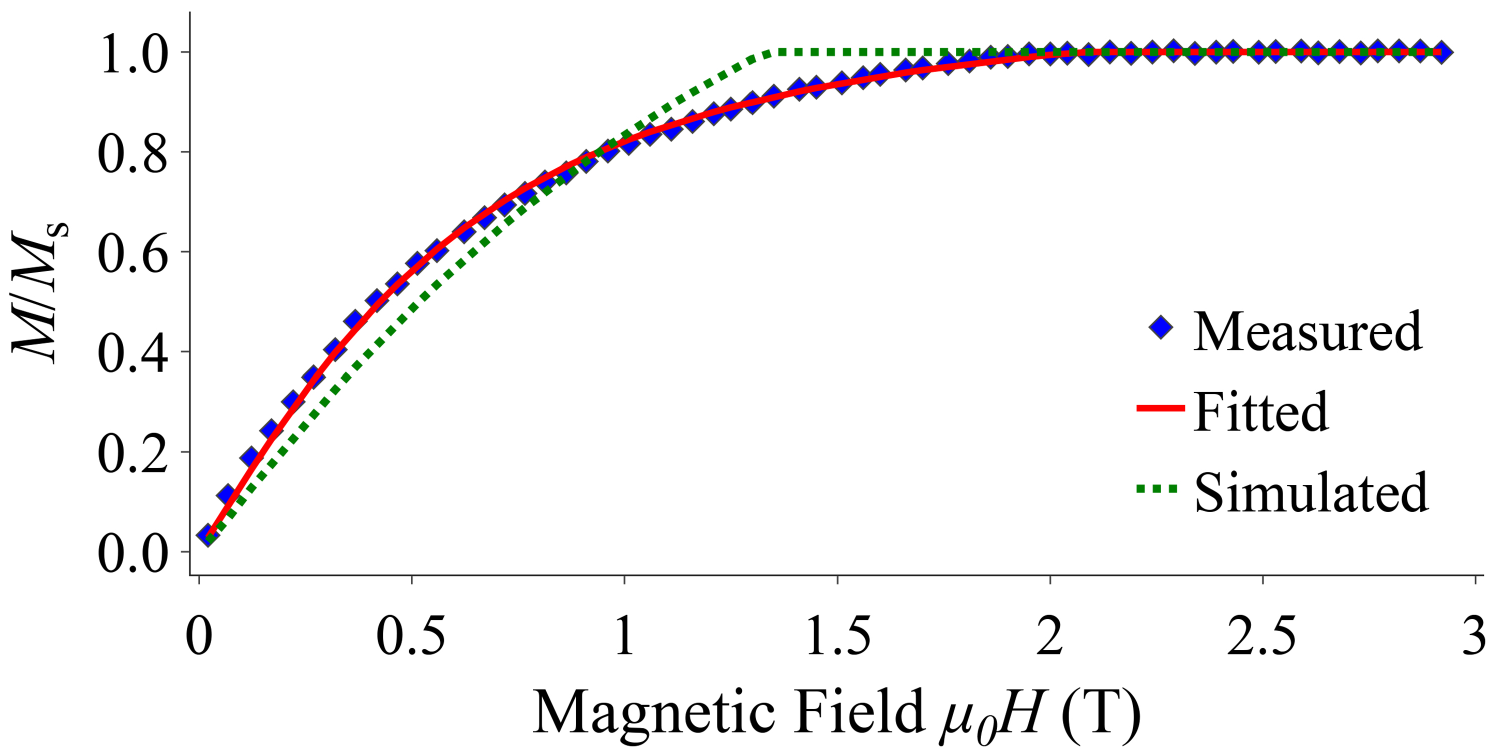}
    \caption{Experimentally measured, simulated, and fitted $M(H)$ curves for an antiferromagnetically coupled Co(5)/Ru(0.4)/Co(5) multilayer (thicknesses in nanometers). The simulation uses a stiffness an order of magnitude larger than that of Co while the fit yields $A_\mathrm{ex} = \SI{15.7 \pm 0.2}{\pico\joule\per\meter}$. Both the fitted and simulated curves assume $J_1 = \SI{3.952 \pm 0.004}{\milli\joule\per\square\meter}$ and $J_2 = \SI{0.251 \pm 0.009}{\milli\joule\per\square\meter}$.}
    \label{fig: infinite_stiffness_fit}
\end{figure}

A commonly used, indirect approach for extracting $J_1$ and $J_2$ from experimental $M(H)$ curves assumes a macrospin model where the two FM layers are infinitely stiff ($A_\mathrm{ex} = \infty$). While capable of capturing the general trend, this approach often fails to accurately determine values for $J_1$ and $J_2$. This is evident from the experimentally measured, simulated, and fitted $M(H)$ curves for Co/Ru/Co shown in \cref{fig: infinite_stiffness_fit}. The simulation is performed with an  exchange constant, $A_\mathrm{ex}$, an order of magnitude larger than that of the Co layers, while the fit yields $A_\mathrm{ex} = \SI{15.7 \pm 0.2}{\pico\joule\per\meter}$. For both the simulated and fitted models, $J_1 = \SI{3.952 \pm 0.004}{\milli\joule\per\square\meter}$ and $J_2 = \SI{0.251 \pm 0.009}{\milli\joule\per\square\meter}$. The simulated data exhibit a pronounced corner discontinuity, whereas the measured and fitted results smoothly approach the saturation magnetization, $M_s$. Thus, to measure the strength of bilinear and biquadratic interlayer exchange coupling using $M(H)$ data from magnetic films with in-plane magnetic anisotropy, one \textit{must} simultaneously fit $J_1$, $J_2$, and $A_\mathrm{ex}$. 

The direct exchange interaction, as quantified by $A_\mathrm{ex}$, is an essential parameter to model magnetization state and reversal in all magnetic materials \cite{abert_micromagnetics_2019}. The most common methods to directly determine $A_\mathrm{ex}$ are through neutron scattering \cite{alperin_observation_1966, shirane_spin_1968, michels_measuring_2000}, Brillouin light scattering (BLS) \cite{vernon_brillouin_1984-1, liu_exchange_1996}, or ferromagnetic resonance (FMR) measurements of spin-wave-dispersion curves. However, the excitation frequency of magnon modes is inversely proportional to the square of the film thickness. For \SI{10}{\nano\meter} Co films, similar to those in this study, this corresponds to a frequency of over $\SI{400}{\giga\hertz}$ \cite{waring_exchange_2023}, which is far higher than the largest microwave frequency used in FMR measurements \cite{eyrich_effects_2014}. As well, neutron scattering measurements of thin films are limited by the large penetration depth of neutrons. Lastly, while BLS measurements can be used to measure stiffness in these thin films, they require an experimental setup that is uncommon in magnetics laboratories \cite{eyrich_effects_2014}. 

Our research group previously published two articles presenting models that simulate magnetization reversal as a function of the applied external field, $M(H)$, in interlayer exchange-coupled thin films. These models either calculate the total magnetic energy in each of a number of discrete atomic sub-layers \cite{girt_method_2011} or assume continuous variation of the magnetic torque across the magnetic layers \cite{omelchenko_continuous_2022}. In both articles, these approaches are used to simulate magnetization curves of FM/SL/FM multilayers, where the FM layers have identical thickness, composition, and layering. The energy model was originally expanded to include biquadratic term by \citet{nunn_control_2020}, and was later extended to asymmetric structures by \citet{mckinnon_thermally_2022}, although no explanation of method was offered. 

This article provides generalizations of our previous theoretical work \cite{girt_method_2011, omelchenko_continuous_2022} to fully model multilayers of arbitrary thicknesses, compositions, and layering complexities, and demonstrates the use of these expanded models by fitting to measured $M(H)$ data of these structures. This capability is important for the characterization and design of the complex thin films commonly used in devices. This approach enables one to design multilayer structures that provide accurate measurements of stiffness in arbitrary ferromagnetic layers, including those with low or no interlayer coupling. Furthermore, this work introduces and validates an entirely novel approach to modeling and fitting to $M(H)$ data, the discrete torque model, which considers the magnetic torque acting on individual atomic layers. We show that this method can be mathematically reduced to the continuous case, but offers substantially improved computational performance and better agreement with the energy model for complex structures. We provide a public-access website where users can fit any $M(H)$ data of noncollinear and antiferromagnetic IEC multilayers with the energy model developed in this paper. 

In this article, we will begin with a detailed derivation of the discrete energy (\cref{sec: discrete energy}), discrete torque 
(\cref{sec: disc torque}), and continuous torque (\cref{sec: cont torque}) models. We then discuss the methods by which these models are fit to experimental data (\cref{sec: methods theoretical}), and the methods use to collect these data (\cref{sec: methods experimental}). We will then present results comparing the statistical and computational performance of the three models when fitting simulated (\cref{sec: results theoretical}) and experimentally measured (\cref{sec: results experimental}) $M(H)$ curves. These data will be compared with previously fitted and directly measured values of the fit parameters across various approaches (\cref{sec: results variable exch,sec: results exch free}). Finally, we will discuss the implications of these novel approaches to fitting $M(H)$ curves for future characterization and device design work (\cref{Sec: Discussion}). 

%%%%%%%%%%%%%%%%%%%%%%%%%%%%%%%%%%%%%%%%%%%%%%%%%%%%%%%%%%%%%%
% THEORY
%%%%%%%%%%%%%%%%%%%%%%%%%%%%%%%%%%%%%%%%%%%%%%%%%%%%%%%%%%%%%%

\section{\label{sec: Theory}Theory}

\begin{figure}
    \subfloat[]{\label{fig:first}%
      \includegraphics[width=0.95\columnwidth]{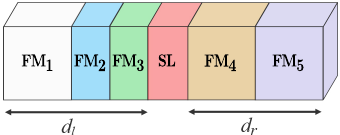}%
    }
    \\[-0.5ex]
    
    \subfloat[]{\label{fig:second}%
      \includegraphics[width=0.95\columnwidth]{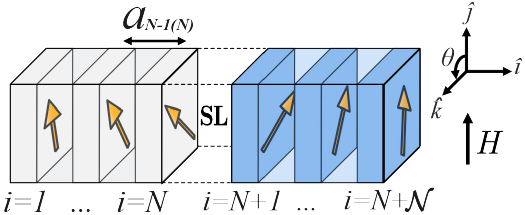}%
    }
    \caption {(a) The magnetic multilayer structure FM$_1$/FM$_{2}$/FM$_{3}$/\allowbreak SL/FM$_{4}$/FM$_5$ and (b) the atomic layer structure used in the discrete energy and discrete torque models. $d_l$ and $d_r$ are the total thickness of ferromagnetic layer on the left and right side of SL, and $a_{N-1(N)}$ the distance between magnetic atomic planes $N-1$ and $N$. Gaps are inserted between the magnetic sub-layers to illustrate the directions of the magnetic moments within each sub-layer.}
    \label{fig: theory diagrams}
\end{figure}

In this article, we develop and refine approaches to model the response of a magnetic multilayer structure to an external magnetic field. The multilayer structure under consideration, FM$_1$/FM$_{2}$/.../SL/.../FM$_{K-1}$/FM$_K$ is illustrated in \Cref{fig: theory diagrams}. It consists of one or more ferromagnetic (FM) layers on either side of a spacer layer (SL) which mediates interlayer exchange coupling between the neighboring FM layers. We assume that the demagnetization fields in the magnetic layers are much larger than any out-of-the-film fields that arise from the surface and magnetocrystalline anisotropies. As a result, the total magnetic anisotropy, and therefore the magnetization of the layers, lies in-plane. We further assume that the samples are polycrystalline and have a random crystal orientation in-the-plane, leading to no preferred direction of magnetization in the plane of the film.

In the models developed in this work, each ferromagnetic layer is discretized into sub-layers. We assume these sub-layer thicknesses to be equal to that of a single atomic layer. We define there to be $N$ magnetic planes in the FM layer to the left of the SL and $\mathcal{N}$ magnetic planes in the FM to the right of the SL. Each plane is assigned a macrospin $\vec{M}_i$ defined by: 
\begin{equation}
    \vec{M}_i = \begin{cases}
     \Ms{i}\left[0, \cos\theta_i, \sin\theta_i\right] & \text{if } i \le N\\
    \Ms{i}\left[0, \cos\theta_i, -\sin\theta_i\right] & \text{if } i > N,
\end{cases}
\end{equation}
where $\Ms{i}$ is the magnetic moment per unit area and $\theta_i$ is the angle of the magnetic moment of the $i^{th}$ atomic plane with respect to an applied external magnetic field, see \Cref{fig: theory diagrams}. The SL is several atomic layers thick and may contain some fraction of magnetic material, but it is considered to have zero magnetic moment in this work. This has been shown to be a good approximation, even in the case where the SL is magnetic \citep{nunn_control_2020}. 

Direct measurements of $A_\mathrm{ex}$ using BLS or neutron scattering are inaccessible to most laboratories, and FMR is not well-suited to approximately sub-\SI{30}{\nano\meter}-thick thin films \cite{eyrich_effects_2014}. Instead, one can use a method based on the formation of an in-plane spin spiral throughout the magnetic atomic planes of FM$_1$/SL/FM$_1$ in an external field, as illustrated in \ref{fig: theory diagrams}(b). Interlayer coupling acts only on the magnetic moments in the atomic layers at the SL interface, aligning them antiferromagnetically or noncollinearly. Meanwhile, the Zeeman effect from the external field acts to align the magnetic moments throughout the FM layers along the field direction. The competition between these effects induces the spin spiral. In FM$_1$/SL/FM$_1$ structures with weak interlayer coupling, direct exchange coupling between FM layers can dominate and form  negligible spin spirals. In such cases, fitting to the $M(H)$ curve cannot accurately determine the stiffness of FM$_1$. However, even in these structures, introducing atomically thin interfacial coupling layers between FM$_1$ and the SL allows for an accurate measurement of $A_\mathrm{ex}$ by fitting to the $M(H)$ curve. 

This work discusses and compares discrete energy, discrete torque, and continuous torque models which are used to fit to the $M(H)$ dependence of magnetic multilayer structures. $M(H)$ is of primary experimental interest as its behavior is easily and directly measured through magnetometry. The discrete energy \citep{girt_method_2011} and continuous torque \citep{omelchenko_continuous_2022} models have been previously developed to simulate simplified, symmetric multilayer structures, FM$_1$/SL/FM$_1$. The energy model was further applied to fit asymmetric structures by \citet{mckinnon_thermally_2022}, however, no explanation of the method was provided. Here, we introduce a new theoretical model, discrete torque, and present mathematical steps to generalize all three models to include multilayers with any number of ferromagnetic layers, differing exchange stiffnesses, layer thicknesses, and saturation magnetizations. These novel extensions offer a methodology for measuring interlayer coupling and stiffness in any antiferromagnetic or noncollinearly-coupled magnetic multilayer structure with in-plane magnetic anisotropy. 

\subsection{Discrete energy}\label{sec: discrete energy}

From the work done by \citet{girt_method_2011}, the total areal magnetic energy density $E_\mathrm{mag}$ of  antiferromagnetically- or noncollinearly-coupled ferromagnetic layers in an external magnetic field is composed of three components:
\begin{equation}
    E_\mathrm{mag} = E_\mathrm{Z} + E_\mathrm{int} + E_\mathrm{ex},
\end{equation}
where $E_{Z}$ is the Zeeman energy, $E_\mathrm{int}$ is the interlayer exchange energy, and $E_\mathrm{ex}$ is the exchange energy. These can be expressed as
\begin{align}
    E_\mathrm{Z} & = -\mu_0H\sum_{i=1}^{N+\mathcal{N}} a_{i} \Ms{i}\cos(\theta_{i}), \label{eq:zeeman energy}\\
    E_\mathrm{ex} & = -2\left( \sum_{i=1}^{N-1} \dfrac{A_{\mathrm{ex},i(i+1)}}{a_{i(i+1)}}\cos(\theta_{i}-\theta_{i+1})  \right. \nonumber\\ & \quad \left.+ \sum_{i=N+1}^{N+\mathcal{N}-1} \dfrac{A_{\mathrm{ex},i(i+1)}}{a_{i(i+1)}} \cos(\theta_{i}-\theta_{i+1})\right), \label{eq:exchange energy}\\
    E_\mathrm{int} & =  J_1 \cos(\theta_{N} - \theta_{N+1}) + J_2 \cos^2 (\theta_{N} - \theta_{N+1}),\label{eq:interlayer energy}
\end{align}
where $H$ is the external field, $\A{i(i+1)}$ and $a_{i(i+1)}$ are the exchange stiffness and interlayer spacing between the $i$th and $i+1$th ferromagnetic atomic planes, $a_{i}$ is the thickness of the $i$th ferromagnetic atomic plane, and $J_1$ and $J_2$ are the bilinear and biquadratic interlayer coupling constants. Note that $a_{i(i+1)}$ and $a_{i}$ are distinct only if the thickness of the $i$th and $i+1$th atomic planes are different. 

If the multilayer structure includes more than one ferromagnetic layer on either side of the SL, \cref{eq:exchange energy} can be adjusted to account for differences in stiffness 
$A_\mathrm{ex}$, atomic lattice spacing 
$a_i$, and the number of atomic layers 
$N_i$ across the various ferromagnetic layers. For instance, for a structure FM$_1$/FM$_2$/SL/FM$_3$, \cref{eq:exchange energy} reduces to 
\begin{equation}
\begin{aligned}
    E_\mathrm{ex} = & -2\left( \dfrac{A_{\mathrm{ex},1}}{a_{1}}\sum_{i=1}^{N_1-1} \cos\left(\theta_{i}-\theta_{i+1}\right) \right. \\ & \left.+   
    \dfrac{A_{\mathrm{ex},2}}{a_{2}}\sum_{i=N_1}^{N-1} \cos\left(\theta_{i}-\theta_{i+1}\right)  \right. \\ & \left.+
    \dfrac{A_{\mathrm{ex},3}}{a_{3}}\sum_{i=N+1}^{N+\mathcal{N}-1} \cos\left(\theta_{i}-\theta_{i+1}\right)\right)
    , \label{eq:asym exchange energy}
\end{aligned}
\end{equation}
where FM$_1$ has $N_1$ atomic layers and FM$_2$ has $N - N_1$ atomic layers, $\A{k}$ is the exchange stiffness, and $a_K$ is the lattice constant in the entirety of FM$_K$. The exchange stiffness and interlayer spacing between the last atomic layer of FM$_1$ and the first atomic layer of FM$_2$ are assumed to be equal to those of FM$_1$ in accordance with the conclusions of \cref{app: transition stiffness}, which shows that the $M(H)$ curve is independent of the choice of transition stiffness, see also \cref{sec: methods theoretical}. 

To use this model to fit to $M(H)$ curves, $\theta_i\left(H\right)$ as a function of $J_1$, $J_2$, $A_{\mathrm{ex},i}$ $a_i$, $a_{i(i+1)}$, and $\Ms{i}$ is  determined by minimizing the total energy with respect to $\theta_i$ at each external field value $\left(\partial E_{mag}/\partial \theta_i|_H = 0\right)$. $J_1$, $J_2$, and $A_{\mathrm{ex},i}$ are the fitting parameters and it is assumed that $a_i$, $a_{i(i+1)}$, and $\Ms{i}$ are known. This dependence is then used to fit the total magnetization along the external magnetic field direction as a function of the field strength, as follows:
\begin{equation}
    M(H) = \dfrac{1}{M_\mathrm{s,tot}}\sum_{i=1}^{N+\mathcal{N}}\Ms{i}\cos\left(\theta_i\left(H\right)\right),
    \label{eq: M(H)}
\end{equation}
where $M_\mathrm{s,tot}$ is the saturation magnetization of the entire multilayer structure.  

\subsection{Discrete torque}\label{sec: disc torque}

The discrete torque model is derived by calculating the total magnetic torque on each ferromagnetic atomic layer of the multilayer structure (in \cref{fig: theory diagrams}) induced by internal and external magnetic fields. The driving principle of this model is that, at equilibrium, the total magnetic torque must equal zero. 

Two torques act on all ferromagnetic atomic layers in the presence of the external magnetic field [which is assumed to be along the $\uvecj$-direction $\vec{H} = H\uvecj$, see \cref{fig: theory diagrams}(b)]: the exchange torque and the field torque. The exchange torque between any adjacent ferromagnetic atomic layers $i$ and $i+1$ is given as (for derivation see \cref{app: torque})
\begin{equation}
\begin{aligned}
     \vec{\tau}_{\mathrm{ex,}(i+1)\to i} &= 2\dfrac{\A{i(i+1)}}{a_{i(i+1)}}\dfrac{\M{i} \times \M{i+1}}{\Ms{i}\Ms{i+1}}, \\
     &= \begin{cases}
         -\dfrac{2\A{i(i+1)}}{a_{i(i+1)}} \sin(\theta_i - \theta_{i+1})\uveci & \text{if } i \le N\\
        \dfrac{2\A{i(i+1)}}{a_{i(i+1)}} \sin(\theta_i - \theta_{i+1})\uveci & \text{if } i > N,
    \end{cases}
    \label{eq: exchange T}
\end{aligned}
\end{equation}
and the external field torque on a ferromagnetic atomic plane $i$ is
\begin{equation}
\begin{aligned}
    \vec{\tau}_{\mathrm{ext},i} &= a_i\M{i}\times \mu_0\vec{H}, \\
    &= \begin{cases}
         -a_{i}\mu_0\Ms{i}H\sin(\theta_i)\uveci & \text{if } i \le N\\
        a_{i}\mu_0\Ms{i}H\sin(\theta_i)\uveci & \text{if } i > N.
    \end{cases}
    \label{eq: external T}
\end{aligned}
\end{equation}
Additionally, the ferromagnetic atomic layers \textit{at the interface} with the spacer layer also experience interlayer exchange torque. Thus, for the layers $i=N$ and $i=N+1$:
\begin{align}
    \vec{\tau}_{\mathrm{int},N}&= -\vec{\tau}_{\mathrm{int},N+1} \nonumber \\&=\left[J_1 \sin\left(\theta_N + \theta_{N+1}\right) + J_2 \sin\left(2\left(\theta_N + \theta_{N+1}\right)\right)\right]\uveci.
    \label{eq: int T}
\end{align}

The net magnetic torque experienced by a given ferromagnetic atomic layer within the structure depends on its location. The interfaces of the structure act as boundary conditions and must be handled separately. For a layer $i$ within the bulk of the ferromagnet, there is exchange torque from the atomic layer to the left $(i-1)$, exchange torque from the atomic layer to the right $(i+1)$, and the external field torque, producing a total torque at equilibrium
\begin{equation}
    \vec{\tau}_{\mathrm{net},i} =  \vec{\tau}_{\mathrm{ex,}(i-1)\to i} + \vec{\tau}_{\mathrm{ex,}(i+1)\to i} +
    \vec{\tau}_{\mathrm{ext},i} = 0.
\end{equation}
These torques are all in the $\uveci$-direction and so, using \cref{eq: exchange T,eq: external T}, the net torque may be written as
\begin{equation}
\begin{aligned}
    & \tau_{\mathrm{net},i}  = \dfrac{2\A{i(i-1)}}{a_{i(i-1)}} \sin(\theta_i - \theta_{i-1})  \\& +  \dfrac{2\A{i(i+1)}}{a_{i(i+1)}} \sin(\theta_i - \theta_{i+1})  + a_{i}\mu_0\Ms{i}H\sin(\theta_i) = 0,
    \label{eq: net torque bulk}
\end{aligned}
\end{equation}
which can be rearranged to derive the following update rule:
\begin{equation}
\begin{aligned}
    \sin(\theta_i-\theta_{i+1}) = &-\dfrac{a_{i(i+1)}\A{i(i-1)}}{a_{i(i-1)}\A{i(i+1)}}\sin(\theta_i - \theta_{i-1}) \\ &- \dfrac{a_{i}a_{i(i+1)}\mu_0\Ms{i}H}{2\A{i(i+1)}}\sin\theta_i.
    \label{eq: T bulk}
\end{aligned}
\end{equation}
This equation applies in the bulk of all ferromagnetic layers in the structure. 

The outermost layers at $i=1$ and $i=N+\mathcal{N}$ have only one neighboring layer and, therefore, the  exchange torque on these layers has only one contribution, from layer 2 and $N+\mathcal{N}-1$, respectively. For the ferromagnetic atomic plane $i=1$, at equilibrium we can write:
\begin{equation}
    \vec{\tau}_{\mathrm{net},1} = \vec{\tau}_{\mathrm{ext},1} + \vec{\tau}_\mathrm{ex,2\to 1} = 0.
\end{equation}
Substituting \cref{eq: exchange T,eq: external T}, we obtain
\begin{equation}
\begin{aligned}
    \vec{\tau}_{\mathrm{net},1}
     = & \dfrac{2\A{1(2)}}{a_{1(2)}}\sin(\theta_1 - \theta_{2}) \\ & +  a_{1(2)}\mu_0\Ms{1}H\sin\theta_1 = 0
\end{aligned}
\end{equation}
and rearranging gives
\begin{equation}
\begin{aligned}
    \sin(\theta_1 - \theta_2) = -\dfrac{a_{1(2)}^2\mu_0\Ms{1}H}{2\A{1(2)}}\sin\theta_1.
    \label{eq: T i=1}
\end{aligned}
\end{equation}
For the magnetic atomic layer $i=N+\mathcal{N}$, the derivation is the same as for $i=1$ and leads to
\begin{equation}
\begin{aligned}
    & \sin(\theta_{N+\mathcal{N}} - \theta_{N+\mathcal{N}-1}) =\\& -\dfrac{a_{N+\mathcal{N}(N+\mathcal{N}-1)}^2\mu_0\Ms{N+\mathcal{N}}H}{2\A{N+\mathcal{N}(N+\mathcal{N}-1)}}\sin\theta_{N+\mathcal{N}}
    \label{eq: T i=t}.
\end{aligned}
\end{equation}

Likewise, there is only one exchange torque acting on the ferromagnetic atomic layers at the interfaces with the spacer, i.e. $i=N$ and $i=N+1$. These originate from the $i=N-1$ and $i=N+2$ layers, respectively. Considering the IEC and external field torques, the total torque for the magnetic layer $i=N$ at equilibrium is
\begin{equation}
    \vec{\tau}_{\mathrm{net},N} = \vec{\tau}_{\mathrm{ext},N} + \vec{\tau}_{\mathrm{ex},(N-1)\to N} + \vec{\tau}_{\mathrm{int},N} = 0.
\end{equation}
Using \cref{eq: exchange T,eq: external T,eq: int T} we obtain
\begin{align}
    0 = &\dfrac{2\A{N(N-1)}}{a_{N(N-1)}}\sin(\theta_N - \theta_{N-1}) -J_1 \sin(\theta_N + \theta_{N+1}) \nonumber \\ & - J_2 \sin(2(\theta_N + \theta_{N+1}))\ + a_N\mu_0\Ms{N}H\sin\theta_N.
    \label{eq: T i=N}
\end{align}
Similarly, equating the total torque on the ferromagnetic atomic layer $i=N+1$ to zero yields
\begin{equation}
    \begin{aligned}
        0 = & \dfrac{2\A{(N+1)(N+2)}}{a_{(N+1)(N+2)}}\sin(\theta_{N+1} - \theta_{N+2}) \\ 
        & - J_1\sin(\theta_N + \theta_{N+1}) - J_2\sin(2(\theta_N+\theta_{N+1})) \\
        & + a_{N+1}\mu_0\Ms{N+1}H\sin\theta_{N+1}.
        \label{eq: T i=N+1}
    \end{aligned}
\end{equation}

For each external magnetic field, $H$, the torque equations \cref{eq: T bulk,eq: T i=1,eq: T i=t,eq: T i=N,eq: T i=N+1}, are numerically solved using the methods described in \cref{app: T solving}. Once an appropriate solution is found, the projection of the magnetic moments onto the field direction may be calculated using \cref{eq: M(H)} and then fit to $M(H)$ curves obtained through magnetometry. 

In general, the multilayer structure may contain multiple ferromagnetic layers on either side of the SL, e.g. FM$_1$/FM$_2$/SL/FM$_3$. In this case, we must consider the torques acting on the ferromagnetic atomic layers at the FM$_1$/FM$_2$ interface. The torque on the \textit{last} atomic layer of FM$_1$, $N_1$, is assumed to be equal to that on all other atomic layers in the bulk of FM$_1$, \cref{eq: net torque bulk}. The torque on the \textit{first} layer of FM$_2$, $N_1 + 1$, can be written as
\begin{equation}
\begin{aligned}
    \tau_{\mathrm{net},N_1 + 1} = & \dfrac{2\A{N_1 + 1 (N_1)}}{a_{N_1 
     + 1 (N_1)}} \sin\left(\theta_{N_1 + 1} - \theta_{N_1}\right) \\& + \dfrac{2\A{N_1 + 1 (N_1 + 2)}}{a_{N_1 + 1 (N_1 + 2)}} \sin\left(\theta_{N_1 + 1} - \theta_{N_1 + 2}\right) \\& + a_{N_1 + 1}\mu_0\Ms{i}H\sin\left(\theta_{N_1 + 1}\right) = 0,
\end{aligned}
\end{equation}
where $\A{N_1 + 1 (N_1)}$ is the stiffness of FM$_1$ and $\A{N_1 + 1 (N_1 + 2)}$ is the stiffness of FM$_2$. This assumption is in accordance with the conclusions of \cref{app: transition stiffness}, also see \cref{sec: methods theoretical}. 

\subsection{Continuous torque}\label{sec: cont torque}

As was initially demonstrated by \citet{omelchenko_continuous_2022}, a continuous torque model may be used for symmetric multilayers (FM$_1$/SL/FM$_1$) to vastly speed up and stabilize numerical solving. The continuous approach can be uniquely derived from the discrete model described in \cref{sec: disc torque}, in the limit where $a_i$ and $a_{i(i+1)}$ approach zero, i.e. the continuum limit. This likewise implies that $\theta_i - \theta_{i+1}$ approaches zero. Applying these limits to \cref{eq: T bulk} leads to
\begin{equation}
\begin{aligned}
    \theta_i - \theta_{i+1} = & -\dfrac{a_{i(i+1)}\A{i(i-1)}}{a_{i(i-1)}\A{i(i+1)}}(\theta_i - \theta_{i-1}) \\ & - \dfrac{a_{i}a_{i(i+1)}\mu_0M_sH}{2\A{i(i+1)}}\sin(\theta_i).
\end{aligned}
\end{equation}
Dividing both sides by $a_{i(i+1)}$ and multiplying by $-\A{i(i+1)}$ gives
\begin{equation}
\begin{aligned}
    \A{i(i+1)}\dfrac{\theta_{i+1} - \theta_i}{a_{i(i+1)}} = &\A{i(i-1)}\dfrac{\theta_i - \theta_{i(i-1)}}{a_{i(i-1)}}  \\ &+ \dfrac{a_i\mu_0M_sH}{2}\sin(\theta_i). 
\end{aligned}
\end{equation}
In the limit where $a_{i(i+1)}, a_{i(i-1)} \to 0$, this becomes
\begin{equation}
    \begin{aligned}
    \A{i(i+1)}\left(\dfrac{d\theta}{dx}\right)_{i+1} = &  \A{i(i-1)}\left(\dfrac{d\theta}{dx}\right)_{i-1} \\ &+ \dfrac{a_{i}\mu_0M_sH}{2}\sin(\theta_i).\label{eq: T cont multi}
\end{aligned}
\end{equation}
We use the convention that the $\uveci$, $\uvecj$, and $\uveck$ unit vectors point along the x, y, and z directions, respectively. Note that $\theta(x)$ continuously varies along the $\uveci$-direction. In the simplest case of two ferromagnetic layers of the same composition coupled across a spacer layer, FM$_1$/SL/FM$_1$ (the symmetric case), $\A{i(i+1)}=\A{i(i-1)} = \A{}$ and $a_i=a$. This results in:
\begin{align}
    \left[\left(\dfrac{d\theta}{dx}\right)_{i+1} - \left(\dfrac{d\theta}{dx}\right)_{i-1}\right]/a &= \dfrac{\mu_0M_sH}{2A_\mathrm{ex}}\sin(\theta_i).
    \label{eq: factor out A}
\end{align}
Finally, applying the limit $a\to0$ gives
\begin{align}\label{eq: cont sym diff eq}
    \dfrac{d\theta^2}{dx^2} = \dfrac{\mu_0M_sH}{2A_\mathrm{ex}}\sin(\theta).&
\end{align}
This expression is in agreement with the former result derived by \citet{omelchenko_continuous_2022}.

If, instead, two ferromagnetic layers with \textit{different} compositions are coupled across a spacer layer, FM$_1$/\allowbreak SL/FM$_2$, the angular dependence of magnetization is described by two equations (subscripted by 1 or 2):
\begin{equation}
    \dfrac{d\theta^2_{1(2)}}{dx^2} = \dfrac{\mu_0M_{s,1(2)}H}{2A_{\mathrm{ex},1(2)}}\sin(\theta_{1(2)}),
    \label{eq: CT diff}
\end{equation}
one describing the magnetization dependence on $x$ in FM$_1$ and the other in FM$_2$.

We now extend the model to asymmetric multilayers with arbitrary numbers of layers, FM$_1$/FM$_{2}$/.../\allowbreak SL/\allowbreak .../FM$_{K-1}$/FM$_K$. In this case, the exchange stiffness is not always constant, $\A{i(i+1)} \neq \A{i(i-1)}$, and it cannot be factored out in \cref{eq: factor out A}. Then, in the limit $a\to0$ one can write
\begin{equation}
    \A{1} \dfrac{d\theta}{dx} \bigg\rvert_{x=x_{1,2}^-} = \A{2} \dfrac{d\theta}{dx}\bigg\rvert_{x=x_{1,2}^+},
\end{equation}
where $x_{1,2}$ is the boundary between FM$_1$ and FM$_2$, and superscripts $+$ and $-$ indicate approaching $x_{1,2}$ from the right and left, respectively.

Boundary conditions at the outermost atomic layers and at the interface with the SL are required to solve the differential equation, \cref{eq: CT diff}. Applying the limits $a_{1(2)}\to0$ to \cref{eq: T i=1} and $a_{N(N+1)},a_N\to0$ to \cref{eq: T i=N} leads to:
\begin{align}
    \dfrac{d\theta}{dx} &= 0 \biggr\rvert_{x=-d_{l}} \label{eq: CT 1},\\
    \dfrac{d\theta}{dx} &= 0 \biggr\rvert_{x=d_{r}}, \label{eq: CT t}
\end{align}
where $d_l$ is the thickness of the entire ferromagnetic multilayer left of the SL, and $d_r$ is the thickness of the entire ferromagnetic multilayer right of the SL, as shown in \Cref{fig: theory diagrams}. Thus, \cref{eq: CT 1} and \cref{eq: CT t} are the boundary conditions at the outside edges of the ferromagnetic layers. Applying limits $a_{N(N+1)},a_N\to0$ to \cref{eq: T i=N} and ${a_{N+1(N+2)}},a_{N+1}\to0$ to \cref{eq: T i=N+1} leads to
\begin{align}
    2A_\mathrm{ex}\left(\dfrac{d\theta}{dx}\right)_{N-1} - J_1\sin(\theta_{N-1} + \theta_{N+1}) \nonumber \\- J_2\sin2(\theta_{N-1} + \theta_{N+1}) &= 0\biggr\rvert_{x=0}, \label{eq: CT N}\\
    -2A_\mathrm{ex}\left(\dfrac{d\theta}{dx}
    \right)_{N+1} - J_1\sin(\theta_{N-1} + \theta_{N+1}) \nonumber \\- J_2\sin2(\theta_{N-1} + \theta_{N+1}) &= 0\biggr\rvert_{x=0}, \label{eq: CT N+1}
\end{align}
where $x=0$ is defined as being at the SL, with the approximation that the SL is infinitely thin. 

For a given value of the applied field $H$, the differential equation [\cref{eq: CT diff}] and the boundary conditions [\cref{eq: CT 1,eq: CT t,eq: CT N,eq: CT N+1}] are solved numerically for $\theta_i$. The projection of the magnetic moment onto the field direction is calculated from the distribution of angles using,
\begin{align}
    M(H) &= \dfrac{1}{M_\mathrm{s,tot}}\int_{-d_l}^{d_r} M_s(x)\cos(\theta(x))dx.
\end{align}

%%%%%%%%%%%%%%%%%%%%%%%%%%%%%%%%%%%%%%%%%%%%%%%%%%%%%%%%%%%%%%
% METHODS
%%%%%%%%%%%%%%%%%%%%%%%%%%%%%%%%%%%%%%%%%%%%%%%%%%%%%%%%%%%%%%

\section{\label{Sec: Methods}Methods}

\subsection{Experimental}\label{sec: methods experimental}

To benchmark and verify the models introduced in \cref{sec: Theory}, we fabricated and measured seven magnetic multilayers, described in \Cref{tbl:samples}. Samples A--E and G are symmetric, while sample F is asymmetric. 

All film layers were deposited on oxidized (100) Si wafers using RF magnetron sputtering at room temperature, with an Ar pressure below \SI{0.27}{\pascal} (\SI{2.0}{\milli\torr}). Prior to deposition, the Si substrates were cleaned using a standard RCA SC-1 process. The substrates were then placed in a load lock chamber, evacuated to approximately \SI{6.7e-5}{\pascal} (\SI{5.0e-7}{\torr}), and transferred without breaking vacuum to a process chamber with a base pressure below \SI{6.7e-6}{\pascal} (\SI{5.0e-8}{\torr}).

The films were deposited using four \SI{5.08}{\centi\meter} (2.00 inch) diameter targets of Ta, Ru, Fe, and Co, with a target-to-substrate distance of approximately \SI{20}{\centi\meter}. The substrate holder rotated constantly during deposition to ensure uniformity in the thickness and composition of the deposited films across the substrate surface. All multilayers were grown on Ta(3.5)/Ru(3.5) seed layers to induce a strong $\langle$0001$\rangle$ growth orientation of hexagonal close-packed CoRu, Co, RuFe, and Ru films. The finished structures were capped with Ru(3.5) to protect the magnetic layers from oxidation. Values in parentheses represent layer thicknesses in nanometers. The entire sputter process was computer-controlled.
\begin{table}[htbp]
    \begin{ruledtabular}
    \begin{tabular}{cc}
    \textbf{Sample} & \textbf{Structure} \\
    \hline
    A & Co(4)/SL/Co(4) \\ \hline
    B & $\coninety(4)$/SL/$\coninety(4)$ \\ \hline
    C & $\coeighty$(4)/SL/$\coeighty$(4) \\ \hline
    D & $\coninety(3.2)$/Co(0.8)/SL/Co(0.8)/$\coninety(3.2)$ \\ \hline
    E & $\coeighty (3.2)$/Co(0.8)/SL/Co(0.8)/$\coeighty(3.2)$ \\ \hline
    F & Co(4)/SL/Co(0.8)/$\coeighty(3.2)$ \\ \hline
    G & \multirow{1}{*}{Co(1.6)/$\coeighty$(1.6)/Co(0.8)/SL/Co(0.8)}/ \\[-3pt]  
     & $\coeighty(1.6)$/Co(1.6) \\[-2pt] 
    \end{tabular}
    \end{ruledtabular}
    \caption{\label{tbl:samples}Summary of sputtered sample structures. Numbers in parentheses represent layer thicknesses in nanometers. The spacer layer (SL) is $\SI{0.5}{\nano\meter}$ thick Ru\textsubscript{29}Fe\textsubscript{71} in all studied structures. The total thickness of the FM layers on each side of the spacer is \SI{4}{\nano\meter}. Samples A--E and G are symmetric, while sample F is asymmetric.}
\end{table}

We applied x-ray reflectivity measurements to determine the growth rates of each material, which were used to control the thickness of each layer. The field dependence of the magnetization, $M(H)$, was measured for each structure using a vibrating sample magnetometer (VSM) and a superconducting quantum interference device (SQUID) in magnetic fields of up to \SI{7}{\tesla}, at \SI{298}{\kelvin}. For all $M(H)$ measurements, the magnetic field was applied parallel to the film plane. 

\subsection{Theoretical}\label{sec: methods theoretical}

Fitting to the $M(H)$ curve of a simple symmetric multilayer, FM$_{1}$/SL/FM$_{1}$, with just one FM layer on each side of the spacer, using the models described in \cref{sec: Theory} requires three fitting parameters: $J_1$, $J_2$ and $\A{1}$. All other material parameters in the models can be inferred from the sample growth process or directly measured. The saturation magnetization $\Ms{1}$ is determined directly from the measured $M(H)$ curves and the film thickness is measured directly from x-ray reflectivity. The atomic spacing in the ferromagnetic layers is known and is measured using x-ray diffraction. In the continuous torque model (\cref{sec: cont torque}), the atomic spacing in the ferromagnetic layers is taken to be infinitely small. The models in this work assume coherent reversal of the magnetic moments within each sub-layer. Thus, we only fit to $M(H)$ data from field values above which the magnetic domains are expelled, ensuring uniform magnetization across each layer \cite{nunn_controlling_2023}.

When the ferromagnetic material on either side of the SL is composed of multiple distinct FM layers, the number of fitting parameters increases with each additional FM layer. For example, fitting a multilayer structure with four FM layers, FM$_1$/FM$_2$/SL/FM$_3$/FM$_4$, requires ten fitting parameters: $J_1$, $J_2$, $\A{1}$, $\A{2}$, $\A{3}$, $\A{4}$, $\Ms{1}$, $\Ms{2}$, $\Ms{3}$, and $\Ms{4}$. In this case, the individual $\Ms{K}$ ($K = 1$ to $4$) cannot be directly determined from the measured $M(H)$ curve. Additionally, the exchange stiffnesses \textit{between} neighboring ferromagnetic layers also need to be fitted, adding another two parameters. Fortunately, the thickness of each FM layer and $a_{K}$ are still known from structural measurements. 

To reduce the number of fitting parameters for FM$_1$/\allowbreak FM$_2$/SL/FM$_3$/FM$_4$ structures, we first fit a set of simpler structures, FM$_K$/SL/FM$_K$, with $K = 1$ to $4$. These simpler symmetric structures, composed of the same ferromagnetic layer, FM$_K$, on either side of the spacer, allow us to independently measure $\Ms{K}$ and determine $\A{K}$.

Subsequently, fitting $M(H)$ for FM$_1$/FM$_2$/SL/FM$_3$/\allowbreak FM$_4$ only requires fitting $J_1$, $J_2$, and possibly two more fitting parameters due to the exchange stiffness at the FM$_1$/FM$_2$ and FM$_3$/FM$_4$ interfaces. However, as demonstrated in \cref{app: transition stiffness}, selecting the exchange stiffness of the FM$_K$/FM$_{K+1}$ interfaces to the average of or equal to $\A{K}$ and $\A{K+1}$ does not significantly affect the $\chi^2$ of the $M(H)$ fit or the best-fit values of $J_1$ and $J_2$. Therefore, at the interfaces, we elect to use a corresponding ``transition stiffness'' equal to the stiffness of the layer farther from the SL for the associated fitting in this work, eliminating $\A{K,K+1}$ as a free parameter. 

%%%%%%%%%%%%%%%%%%%%%%%%%%%%%%%%%%%%%%%%%%%%%%%%%%%%%%%%%%%%%%
% RESULTS
%%%%%%%%%%%%%%%%%%%%%%%%%%%%%%%%%%%%%%%%%%%%%%%%%%%%%%%%%%%%%%

\section{\label{Sec: Results}Results}

\subsection{Simulated comparison of theoretical models}\label{sec: results theoretical}

\begin{figure}[htbp]
    \subfloat[]{\label{fig: asymm multilayer diagram}%
      \includegraphics[width=0.95\columnwidth]{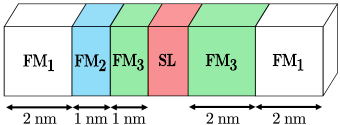}%
    }%
    \\[-0.5ex]
    \subfloat[]{\label{fig: M(H) comparison}%
      \includegraphics[width=\columnwidth]{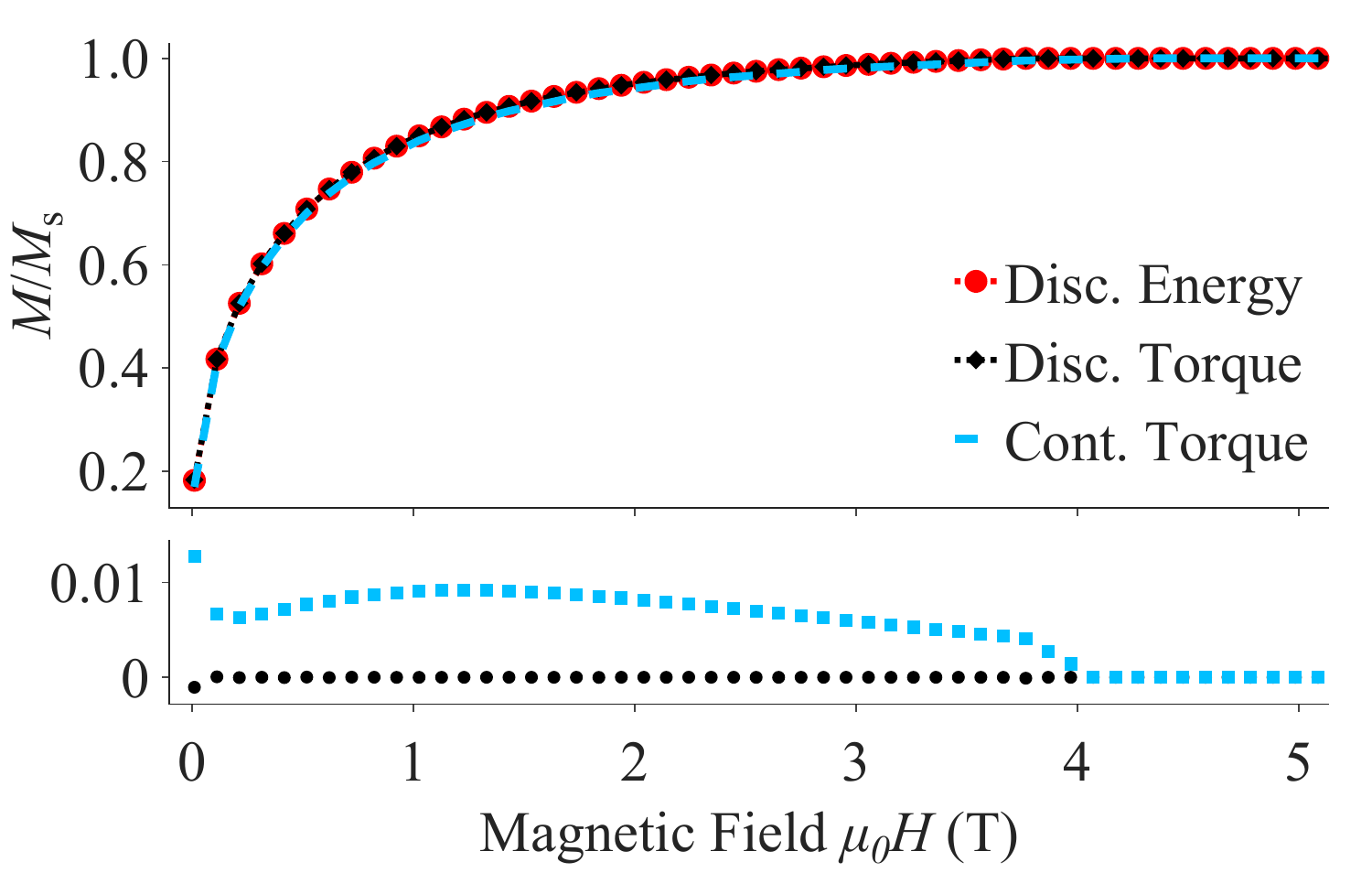}%
    }%
    \\[-0.5ex]
    \subfloat[]{\label{fig: theta comparison}%
      \includegraphics[width=\columnwidth]{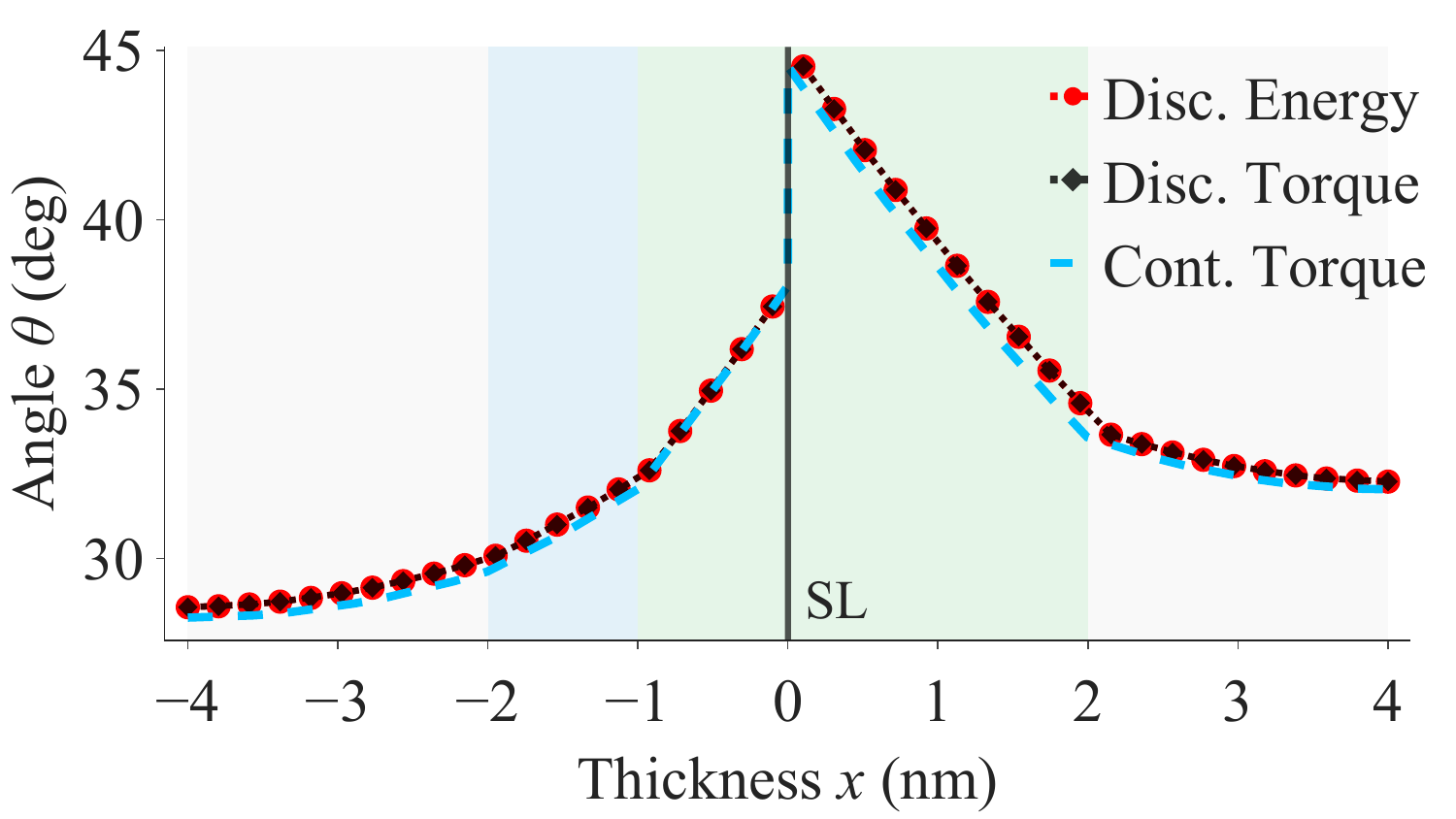}%
    }%
    \caption{
    (a) A schematic diagram of an asymmetric multilayer structure, FM$_1$/FM$_2$/FM$_3$/\allowbreak SL/FM$_3$/FM$_1$, (b) a comparison of the simulated $M(H)$ curves for the asymmetric structure across the three different models, with a sub-figure showing the differences between the two torque models and the energy model, and
    (c) the angular distribution ($\theta_i$) derived from the $M(H)$ curve at an applied field of $\mu_0H = \SI{1}{\tesla}$, with highlighted regions corresponding to the specific modeled structure.
    }
    \label{fig: model comparison}
\end{figure}

To directly compare the three models described in \cref{sec: Theory}, we simulate the $M(H)$ curve of a five-layer asymmetric structure, FM$_1$/FM$_2$/FM$_3$/SL/FM$_3$/FM$_1$, shown in \cref{fig: model comparison}(a). We use the following material parameters in these simulations: $J_1 = \SI{2}{\milli\joule\per\meter\squared}$, $J_2 = \SI{1}{\milli\joule\per\meter\squared}$,  $d_l = d_r = \SI{4}{\nano\meter}$, $\A{1} = \SI{30}{\pico\joule\per\meter}$, $M_{s,1} = \SI{1.5}{\mega\ampere\per\meter}$, $\A{2} = \SI{20}{\pico\joule\per\meter}$, $M_{s,2} = \SI{1}{\mega\ampere\per\meter}$, $\A{3} = \SI{10}{\pico\joule\per\meter}$, and $M_{s,3} = \SI{0.5}{\mega\ampere\per\meter}$. These values are within the range of measurements for pure Co and Co$_x$Ru$_{1-x}$ alloys, as will be shown in this article. All sputtered films are grown in the $\langle$0001$\rangle$ direction, so that the distances between \{0001\} planes are $c/2 = \SI{0.2035}{\nano\meter}$, where $c$ is the lattice constant of the HCP Co structure. We approximate this value to $a_i = \SI{0.2}{\nano\meter}$ for all FM layers.

As shown in \cref{fig: model comparison}(b), the energy and discrete torque models produce nearly identical $M(H)$ curves, while the continuous torque model differs only slightly from the other two. As discussed by \citet{omelchenko_continuous_2022}, the agreement between the energy and continuous torque models improves with the number of magnetic atomic layers. In these simulations, the number of magnetic atomic layers on each side of the spacer is 20, i.e. $N = \mathcal{N} = 20$. 

The angle of each ferromagnetic atomic layer in the simulated structure in a \SI{1}{\tesla} external magnetic field is plotted in \cref{fig: model comparison}(c) for each model. The agreement between the energy and discrete torque models is better than that between the energy and continuous torque models, which is again due to the small number of magnetic atomic layers in the studied structure.

\begin{figure}[htbp]
    \centering
    \includegraphics[width=\columnwidth]{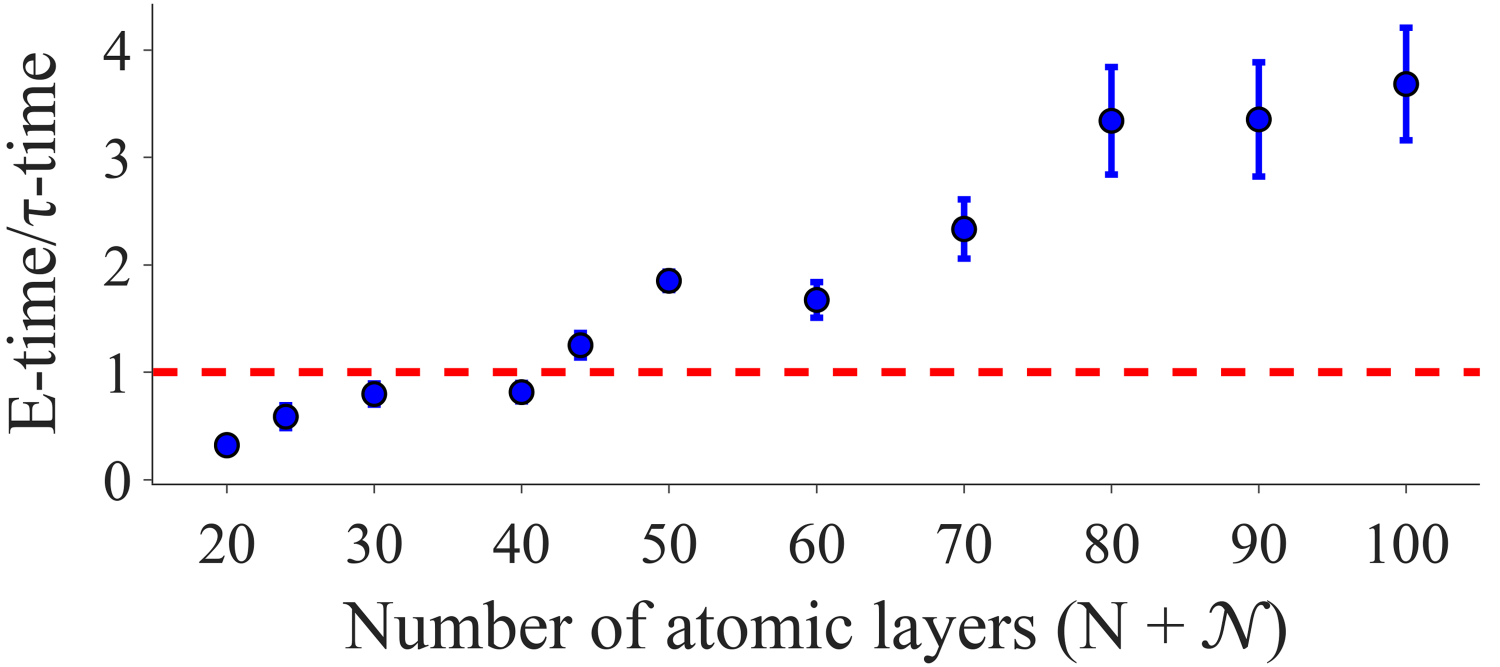}
    \caption{Relative computational performance of the discrete energy and torque models, in terms of the execution time to generate a single $M(H)$ curve of FM$_1$/\allowbreak FM$_2$/\allowbreak FM$_3$/SL/FM$_3$/FM$_1$, as a function of the total number of atomic layers ($N+\mathcal{N}$). The number of atomic layers varies only in FM$_1$, while the thicknesses  of FM$_2$ and FM$_3$ are unchanged. The critical ratio ($E\mathrm{-time}/\lctau\mathrm{-time} = 1$) is indicated by a horizontal line, which is crossed at approximately 42 atomic layers, $N+\mathcal{N} = 42$.}
    \label{fig: performance}
\end{figure}

Figure~\ref{fig: performance} illustrates the relative execution times of the energy and discrete torque models, $E\mathrm{-time}/\lctau\mathrm{-time}$, when generating a single $M(H)$ curve for the FM$_1$/\allowbreak FM$_2$/FM$_3$/SL/FM$_3$/FM$_1$ structure, as a function of $N+\mathcal{N}$. In these simulations, the total number of atomic layers was increased by adding atomic layers of thickness $a_i = \SI{0.2}{\nano\meter}$ to both FM$_1$ layers in \cref{fig: model comparison}(a). $E\mathrm{-time}/\lctau\mathrm{-time}$ exhibits a positive linear trend: for $N+\mathcal{N} > 42$, the discrete torque model outperforms the discrete energy model. The improved scalability of the torque model can be attributed to the update rule [\cref{eq: T bulk}]: increasing $N+\mathcal{N}$ by one requires only one additional computation step. The discrete torque algorithm has a runtime complexity of $\mathcal{O}(N+\mathcal{N})$, i.e. linear in $N+\mathcal{N}$. Conversely, in the energy model, each additional layer increases the dimensionality of the energy minimization problem by one, which contributes to the poorer runtime scaling for large $N + \mathcal{N}$. However, there is additional variable computational overhead associated with rejecting failed solutions to the discrete torque model, so it is outperformed by the energy model for smaller values of $N+\mathcal{N}$. See \cref{app: T solving} for a discussion of the methods used to solve the discrete torque model and a proposal to address this shortcoming in future work. 

The continuous torque model converges faster than the discrete energy and torque models for symmetric single-layered structures. However, its performance suffers greatly when modeling multi-ferromagnetic or asymmetric multilayers. If we assume a total magnetic thickness in \cref{fig: model comparison}(a) of $\SI{8}{\nano\meter}$ and an interlayer spacing of $a=\SI{0.2}{\nano\meter}$ (so $N+\mathcal{N} = 40$), the continuous model is an order of magnitude slower than the discrete torque model. This difference is due to the difficulty of accurately propagating a solution to a differential equation [\cref{eq: CT diff}] through the multilayer, compared to simply iterating 40 steps of a deterministic update rule. The computational performance of the continuous model depends on the total thickness, $d_l + d_r$, rather than $N+\mathcal{N}$. Thus, if the thickness is held constant and $N + \mathcal{N}$ continues to increase, the continuous model would eventually start to outperform the discrete model. However, in practice, it is entirely non-physical to arbitrarily increase $N + \mathcal{N}$ without also increasing the thickness of the FM layers, as this number is determined by the atomic spacing. Thus, the continuous model will always be slower to fit to $M(H)$ data for asymmetric structures than the discrete model. This suggests that a symmetric discrete torque model could also be developed to outperform the symmetric continuous model developed by \citet{omelchenko_continuous_2022}, by avoiding the computational complexities discussed in \cref{app: T solving} and leveraging the update rule.

\subsection{\label{sec: results experimental}Fitting to experimental data}

We used Co, $\coninety$, and $\coeighty$ to construct multilayers consisting of up to six ferromagnetic layers. These materials were chosen because of the large differences in their stiffnesses and magnetizations \citet{eyrich_exchange_2012-1}. A \SI{0.5}{\nano\meter}-thick SL composed of Ru$_{29}$Fe$_{71}$ was used in all studied multilayers.  

\begin{figure}[htbp]
    \centering
    \includegraphics[width=\linewidth]{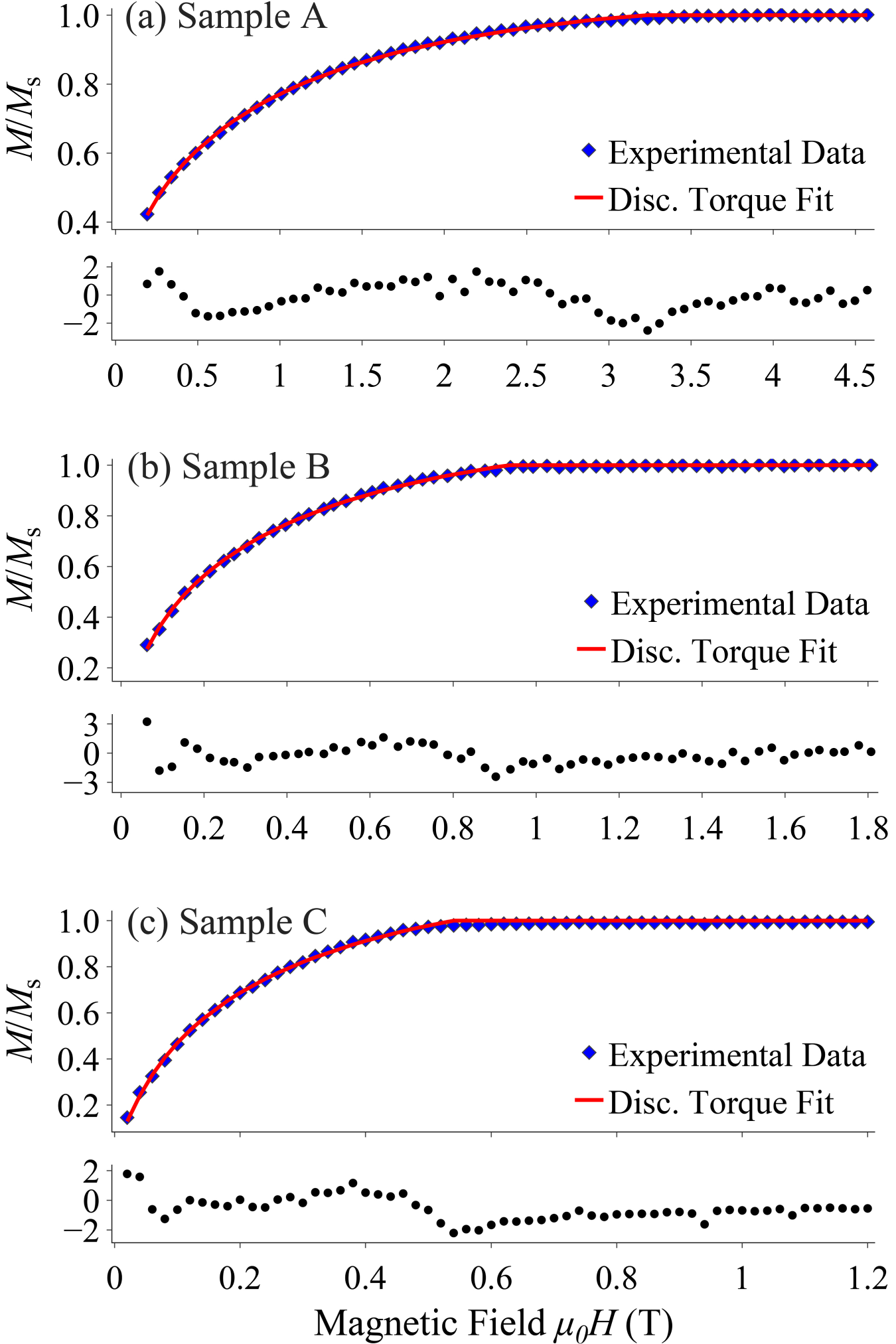}
    \caption{Fitted $M(H)$ curves using the discrete torque model for multilayer structures consisting of only one ferromagnetic material, FM$_1$/SL/FM$_1$, where FM$_1$ is Co (sample A), $\coninety$ (sample B), and $\coeighty$ (sample C), as in \cref{tbl:samples}. Each fit is accompanied by a sub-figure illustrating the standardized residuals. See \Cref{tbl: stiffnesses,tbl: fits} for the fitted model parameters.}
    \label{fig: calibration fits}
\end{figure}

We first study the simple, symmetric structures: Co/SL/Co (sample A), $\coninety$/SL/$\coninety$ (sample B), and $\coeighty$/SL/$\coeighty$ (sample C). Fits with the discrete torque model of a normalized (measured) $M(H)$ curve for each structure are shown in \Cref{fig: calibration fits}. These fits are replicated for the other two models and the best fit values of $A_\mathrm{ex}$ obtained from these fits are reported in \cref{tbl: stiffnesses}. 
\begin{table}[htbp]
    \begin{ruledtabular}
        \begin{tabular}{c|ccc}
            \multirow{2}{*}{\textbf{Sample}} &
            \multicolumn{3}{c}{\boldmath{${A_\mathrm{ex}}$}
            $\mathbf{(\SI{}{\pico\joule\per\meter})}$} \\ 
            & \textbf{Disc. Energy} & \textbf{Disc. Torque} & \textbf{Cont. Torque}\\ \hline
            A & 27.5(5)  & 27.7(5)  & 30.0(5) \\
            B & 13.0(4)  & 12.5(5)  & 13.6(5) \\
            C & 7.2(4) & 7.0(5) & 6.6(4) \\
        \end{tabular}
    \end{ruledtabular}
    \caption{Best fit values of exchange stiffness for the symmetric structures, FM$_1$/SL/FM$_1$, as demonstrated for the discrete torque model in \cref{fig: calibration fits}. FM$_1$ is Co (sample A), $\coninety$ (sample B), or $\coeighty$ (sample C). Values are reported for the discrete energy, discrete torque, and continuous torque models.}
    \label{tbl: stiffnesses}
\end{table}

\begin{figure*}[htbp]
    \centering
    \includegraphics[width=1\linewidth]{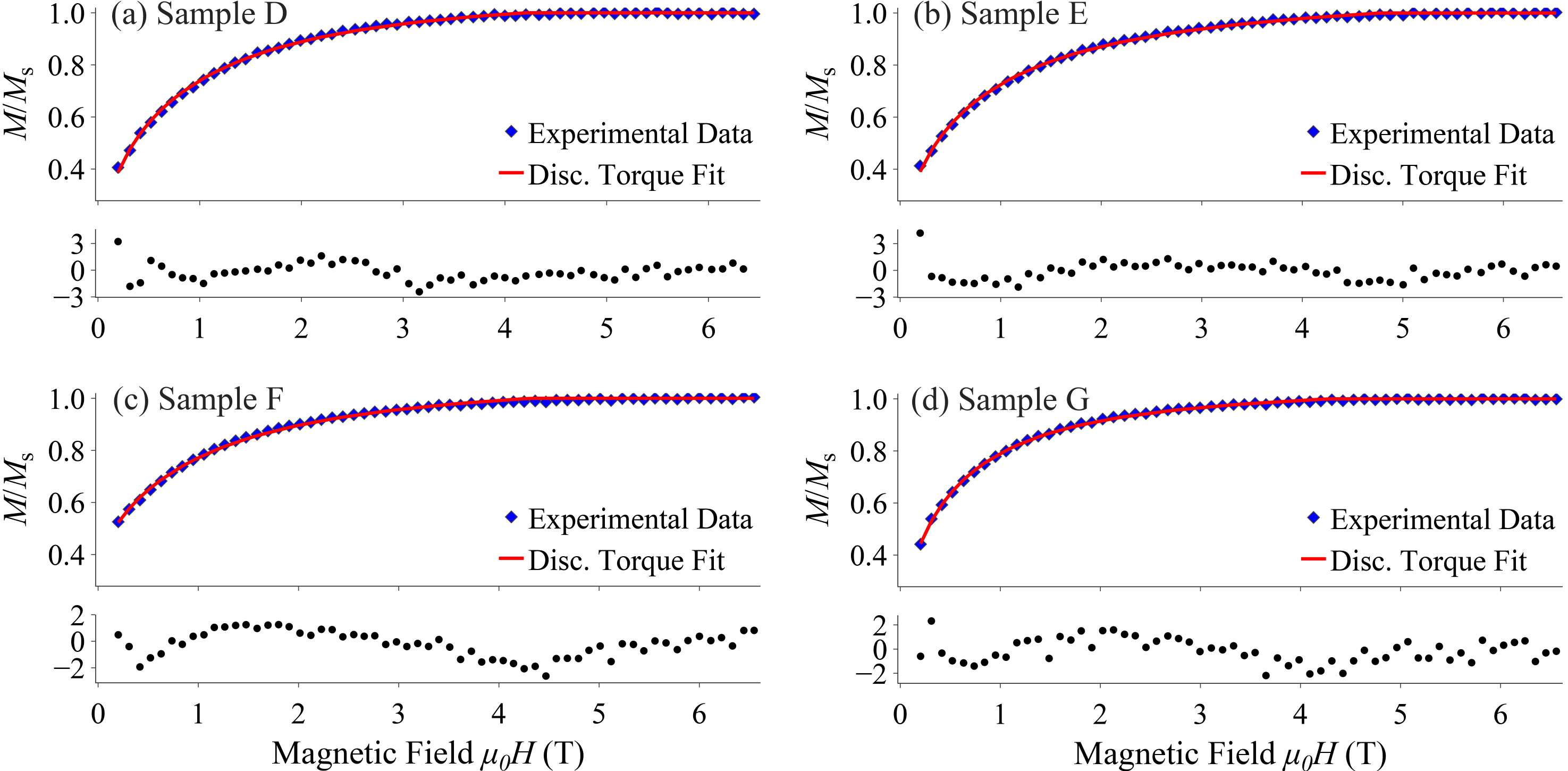}
    \caption{Fits of $M(H)$ curves of the multilayer structures in \cref{tbl:samples} that consist of more than one ferromagnetic material, using the discrete torque model. Below each fit is a sub-figure illustrating the standardized residuals. See samples D, E, F, and G in \Cref{tbl: fits} for the fitted model parameters.}
    \label{fig: M(H) fits}
\end{figure*}

For the subsequent analysis of the multi-ferromagnetic structures (samples D--G), we fit by constraining $A_\mathrm{ex}$ to the values obtained from the energy model for each layer composition, as reported in \cref{tbl: stiffnesses}. Only $J_1$ and $J_2$ remain as free parameters. Figure~\ref{fig: M(H) fits} shows the fits of normalized (measured) $M(H)$ curves for these structures using the discrete torque model. These fits are replicated for the other two models and the best fit values of $J_1$ and $J_2$ for each model obtained from these fits are reported in \Cref{tbl: fits}, along with fit statistics. 
\begin{table*}[htbp]
    \begin{ruledtabular}
    \setlength{\tabcolsep}{4pt}
    \begin{tabular}{c|ccc|ccc|ccc}
         \multirow{2}{*}{\textbf{Sample}} 
         & \multicolumn{3}{c|}{\textbf{Discrete Energy}} 
         & \multicolumn{3}{c|}{\textbf{Discrete Torque}} 
         & \multicolumn{3}{c}{\textbf{Continuous Torque}} \\
         & \multicolumn{1}{c}{\boldmath{$J_1$} $\mathbf{(\SI{}{\milli\joule \per\meter\squared})}$}
         & \multicolumn{1}{c}{\boldmath{$J_2$} $\mathbf{(\SI{}{\milli\joule\per\meter\squared})}$}
         & \multicolumn{1}{c|}{$\bm{\chi^2_\nu}$} 
         & \multicolumn{1}{c}{\boldmath{$J_1$} $\mathbf{(\SI{}{\milli\joule \per\meter\squared})}$}
         & \multicolumn{1}{c}{\boldmath{$J_2$} $\mathbf{(\SI{}{\milli\joule\per\meter\squared})}$}
         & \multicolumn{1}{c|}{$\bm{\chi^2_\nu}$} 
         & \multicolumn{1}{c}{\boldmath{$J_1$} $\mathbf{(\SI{}{\milli\joule \per\meter\squared})}$}
         & \multicolumn{1}{c}{\boldmath{$J_2$} $\mathbf{(\SI{}{\milli\joule\per\meter\squared})}$} 
         & \multicolumn{1}{c}{$\bm{\chi^2_\nu}$} \\ \hline
         A & 3.443(6)   & 1.551(8)  & 4.2 & 3.443(6)   & 1.552(8)  & 4.2 & 3.438(6)  & 1.551(8)    & 4.2 \\
         B & 1.050(2) & 0.314(3) & 3.6 & 1.050(2) & 0.314(3) & 3.6 & 1.049(2) & 0.314(3) & 3.6 \\
         C & 0.487(1) & 0.122(2)  & 6.6 & 0.487(1)  & 0.122(2) & 6.6 & 0.488(1)  & 0.120(2) & 6.5 \\
         D & 3.349(7)   & 1.434(7)   & 5.1 & 3.349(7)   & 1.434(7)   & 5.1 & 3.096(6)   & 1.340(6)   & 6.7 \\
         E & 2.887(7)   & 1.302(7)   & 4.0 & 2.883(6)   & 1.295(7)   & 4.1 & 2.623(6)  & 1.211(6)  & 6.7 \\
         F & 2.851(6)   & 1.761(7)   & 8.0 & 2.851(6)   & 1.761(7)   & 8.0 & 2.224(6)  & 1.867(7)  & 12.2 \\
         G & 2.879(7)   & 1.284(7)   & 2.7 & 2.879(7)   & 1.284(7)   & 2.7 & 2.700(6)   & 1.215(6)   & 5.3 \\
    \end{tabular}
    \end{ruledtabular}
    \caption{\label{tbl:fits} Fitted model parameters from \Cref{fig: calibration fits,fig: M(H) fits} for the discrete energy, discrete torque, and continuous torque models. The corresponding sample structures are given in \Cref{tbl:samples}.}
    \label{tbl: fits}
\end{table*}

The data show that the addition of Ru to Co in Co$_x$Ru$_{1-x}$/SL/Co$_x$Ru$_{1-x}$ significantly reduces the coupling strength between the Co$_x$Ru$_{1-x}$ layers, determined by $J_1$ and $J_2$. The introduction of Ru also reduces both the stiffness and saturation magnetization of pure Co layers from $\A{\text{Co}} = \SI{27.5(5)}{\pico\joule\per\meter}$ and $\Ms{\text{Co}} = \SI{1400 \pm 14}{\kilo\ampere\per\meter}$, to $\A{\coninety} = \SI{13.0(4)}{\pico\joule\per\meter}$ and  $\Ms{\coninety} = \SI{1070 \pm 11}{\kilo\ampere\per\meter}$, and $\A{\coeighty} = \SI{7.2(4)}{\pico\joule\per\meter}$ and  $\Ms{\coeighty} = \SI{774 \pm 8}{\kilo\ampere\per\meter}$.
The observed reduction of $M_\text{s}$ of CoRu alloys is in agreement with the results obtained by \citet{eyrich_effects_2014}. However, the obtained stiffnesses of Co and $\coninety$ alloys are significantly larger ($\sim 2$ times) than previously observed.

All three models yield similar best-fit values of $J_1$, $J_2$, and $A_\mathrm{ex}$ when considering the simple symmetric FM$_1$/SL/FM$_1$ structures, samples A--D, \cref{tbl: fits}. Consequently, the quality of the fits, as determined by $\chi^2_\nu$, are nearly identical. For the more complicated multi-ferromagnetic structures, the discrete models are in excellent agreement, whereas the continuous approximation differs significantly, with the difference increasing with sample structure complexity. Furthermore, the quality of the fit for structures E, F, and G is significantly worse when using the continuous torque model compared to the discrete energy and torque models. This suggests that the continuous torque struggles to find $J_1$ and $J_2$ values that result in a minimized $\chi^2_\nu$. There exist discrepancies in the best-fit values of $J_1$ and $J_2$ between the discrete and continuous models for samples D--G. This is a consequence of the fact that the continuous model assumes a smooth and gradual change in angle $\theta$, which is an approximation that fails for films with low thicknesses (a few atomic layers) and many discontinuities in the $A_\mathrm{ex}$ and $M_s$ distributions, as can be seen in \cref{fig: model comparison}(c). 

In the reported best-fit values, we assume that there is no uncertainty originating from the reproducibility of sample deposition or measurement: the  uncertainties are assumed to arise solely from the fitting process of $M(H)$, where each experimental point carries the uncertainty of the magnetometer. This simplification allows for direct comparison of fitting performance and agreement between the models for identical measured data. 

With the addition of 15 at. \% Ru into the magnetic Co layer the coupling decreases by nearly an order of magnitude, which is in agreement with the extrapolated trend from the previous experimental work \citep{eyrich_effects_2014}. To alleviate this effect, one can incorporate a thin, strongly coupled interface layer on either side of the SL, such as pure Co. This approach allows for strong bilinear and biquadratic coupling even across Co$_x$Ru$_{1-x}$ FM layers, as shown in \Cref{tbl:fits} for samples A and D--G, while maintaining control over all other magnetic properties in the bulk of the layer. This provides far greater freedom in the design of coupled devices, as well as enabling the measurement of stiffness in a wider range of magnetic materials.

\subsubsection{\label{sec: results variable exch}Modeling the Co/Ru interface}

\begin{figure}[htbp]
    \centering
    \includegraphics[width=\columnwidth]{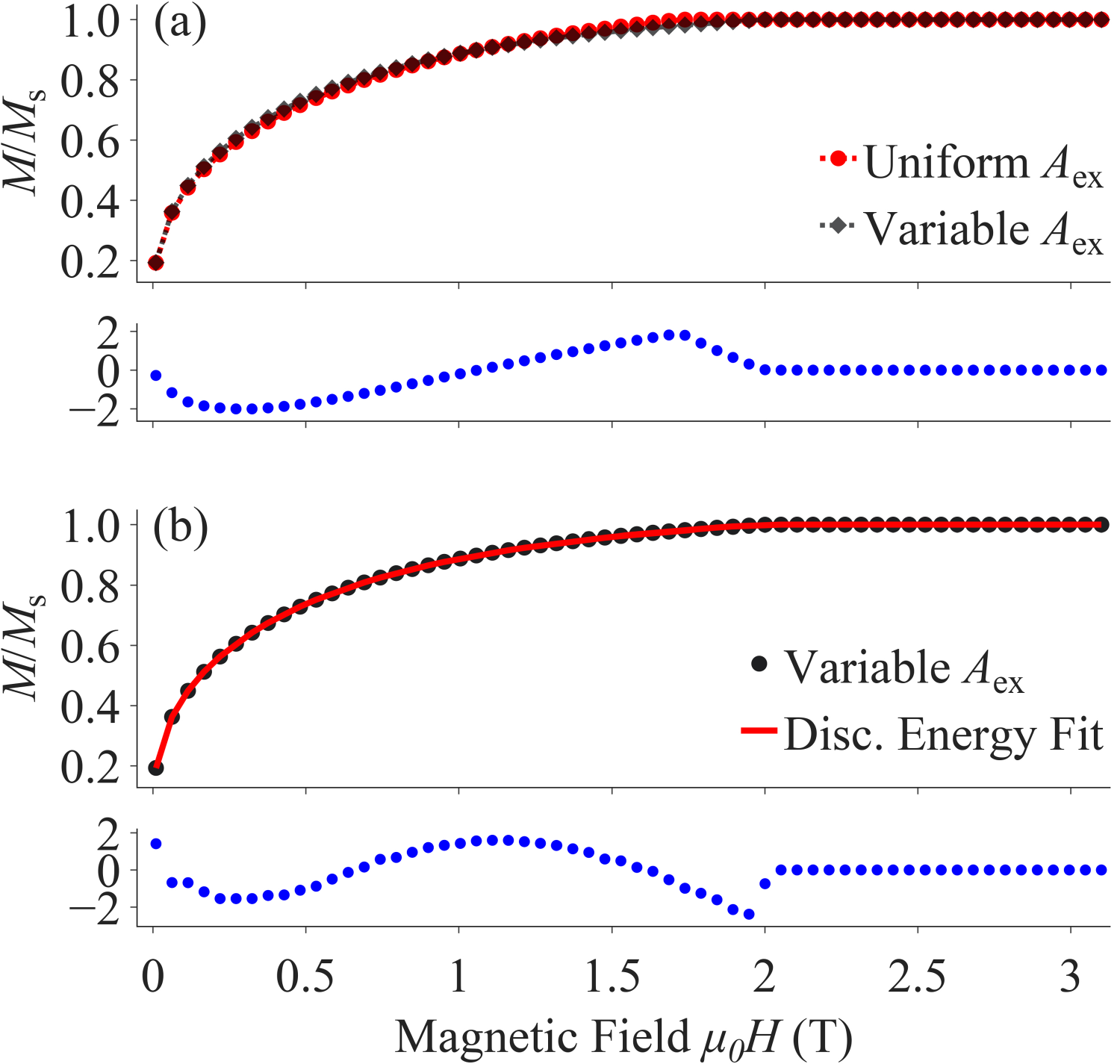}
    \caption{(a) Comparison of uniform and variable exchange stiffness models. In the uniform model, all layers have an exchange stiffness of \SI{27.5}{\pico\joule\per\meter}. The variable model reduces the exchange stiffness to \SI{8}{\pico\joule\per\meter} for two Co atomic layers at the interfaces of the SL, while maintaining an exchange stiffness of \SI{27.5}{\pico\joule\per\meter} for the other layers. The data were generated using the energy model with $J_1 = \SI{2}{\milli\joule\per\square\meter}$ and $J_2 = \SI{1}{\milli\joule\per\square\meter}$. Below the fit is a plot of the percentage difference between the uniform and variable stiffness models. (b) A fit of the energy model, assuming homogeneous stiffness, to an $M(H)$ curve generated from the variable stiffness model, yielding $A_\mathrm{ex} = \SI{16.03 \pm 0.03}{\pico\joule\per\meter}$. Below the fit is a plot of the standardized residuals.}
    \label{fig: uniform vs variable}
\end{figure}

The $A_\mathrm{ex}$ of Co obtained from fitting the energy model to the $M(H)$ of structure A, Co/Ru$_{29}$Fe$_{71}$/Co, is twice as large as the $A_\mathrm{ex}$ of Co obtained from fitting a pure Ru SL structure, Co/Ru/Co \citep{eyrich_effects_2014}. To explain this discrepancy, we simulate the $M(H)$ curve of Co/Ru/Co using the energy model. The simulation is such that each Co layer consists of 20 atomic layers and that the two atomic layers on each side of the Ru SL have lower stiffness than the rest: $A_\mathrm{ex} = \SI{8}{\pico\joule\per\meter}$ at the interfaces and \SI{27.5}{\pico\joule\per\meter} in the bulk of the layer. The bulk stiffness value is that obtained for Co in \cref{tbl: stiffnesses} and we choose the value of stiffness for Co$_{85}$Ru$_{15}$ to produce good agreement with \citet{eyrich_effects_2014}. while being similar to value obtained for Co$_{85}$Ru$_{15}$ in \cref{tbl: stiffnesses}. This modification of the substructure of the Co layer is justified by our previous observations and modeling \citep{eyrich_effects_2014}, which show that atomic layers of Co adjacent to a Ru layer are expected to have lower stiffness. We compare this curve to one assuming a homogeneous value of $A_\mathrm{ex}$ throughout the layers in \cref{fig: uniform vs variable}(a). The coupling constants are taken to be $J_1 = \SI{2.0}{\milli\joule\per\square\meter}$ and $J_2 = \SI{1.0}{\milli\joule\per\square\meter}$ in both simulations. 

Next, we fit the energy model to this simulated $M(H)$, but now assume a uniform stiffness throughout the Co layers. This fit, shown in \cref{fig: uniform vs variable}(b), indicates that a bulk Co stiffness of \SI{16.0 \pm 0.1}{\pico\joule\per\meter} is optimal to characterize the simulated data. This is close to the value reported by \citet{eyrich_effects_2014}: \SI{15.5 \pm 0.1}{\pico\joule\per\meter}. The fit also gives coupling values of $J_1 = \SI{2.000 \pm 0.003}{\milli\joule\per\square\meter}$ and $J_2 = \SI{1.001 \pm 0.003}{\milli\joule\per\square\meter}$, which are equal to those used to generate the simulated data to within the standard error. Thus, the reduction in the stiffness of the atomic layers at the SL interface will strongly affect the fitted value of $A_\mathrm{ex}$ if the model assumes a uniform stiffness. However, a reduction in the interfacial stiffness will not significantly impact the fitted values of $J_1$ and $J_2$.

Alternatively, the fitted value of $A_\mathrm{ex}$ saturates when increasing the stiffness of the interfacial Co atomic layers. When we double the $A_\mathrm{ex}$ of the two layers on each side of the SL to \SI{55.0}{\pico\joule\per\meter}, we measure an overall stiffness of $A_\mathrm{ex} = \SI{32.1 \pm 0.1}{\pico\joule\per\meter}$, an increase of only \SI{4.6 \pm 0.1}{\pico\joule\per\meter}. 

In the Co/Ru$_{29}$Fe$_{71}$/Co structure (sample A), most of the Ru atoms in the SL are replaced with Fe. The presence of Fe is known to not significantly affect the stiffness of Co layers \citep{eyrich_effects_2014}. Consequently, the Ru$_{29}$Fe$_{71}$ SL has a significantly reduced impact on the Co atomic layers at the interface as compared to a pure Ru SL, resulting in a much higher stiffness of Co determined by fitting to the $M(H)$ of Co/Ru$_{29}$Fe$_{71}$/Co, assuming a uniform stiffness across the Co layers. The value of $A_\mathrm{ex} = \SI{27.5 \pm 0.5}{\pico\joule\per\meter}$, obtained from fitting the $M(H)$ of Co/Ru$_{29}$Fe$_{71}$/Co is in much better agreement with previous Brillouin ($A_\mathrm{ex} = \qtyrange{16}{28}{\pico\joule\per\meter}$ \citep{vernon_brillouin_1984-1, liu_exchange_1996}) and neutron ($A_\mathrm{ex} = \qtyrange{27}{35}{\pico\joule\per\meter}$ \citep{alperin_observation_1966, shirane_spin_1968, michels_measuring_2000}) scattering measurements. 

All sputtered magnetic structures in this study are grown on top of a Ru seed layer and covered by a Ru capping layer. These bounding layers may have a similar effect as the Ru at the SL on the measured value of $A_\mathrm{ex}$ for Co, so we also simulate the $M(H)$ curve of a Ru/Co/Ru/Co/Ru structure, where each Co layer consists of 20 atomic layers. Similarly to the above, the stiffnesses of the two atomic Co layers adjacent to the Ru SL and the two atomic Co layers adjacent to the top and bottom Ru layers are lower than that of the remaining 16 Co atomic layers, $A_\mathrm{ex} = \SI{8.0}{\pico\joule\per\meter}$ and \SI{27.5}{\pico\joule\per\meter}, respectively. The simulated $M(H)$ curve for this structure is nearly identical to that with layers of reduced stiffness only at the SL interface. $A_\mathrm{ex}$ remains the same: \SI{16.03 \pm 0.03}{\pico\joule\per\meter} to \SI{15.99 \pm 0.03}{\pico\joule\per\meter}. This is expected, as the reduced stiffness of the Co atomic layers adjacent to the outer Ru primarily affects the reversal in only these layers. In contrast, the reduced stiffness near the SL interface influences the switching behavior of the entire structure by reducing the ability to form spin spirals throughout the multilayer. 

\subsubsection{\label{sec: results exch free}Exchange as a free parameter}

\begin{figure}[htbp]
    \centering
    \includegraphics[width=\columnwidth]{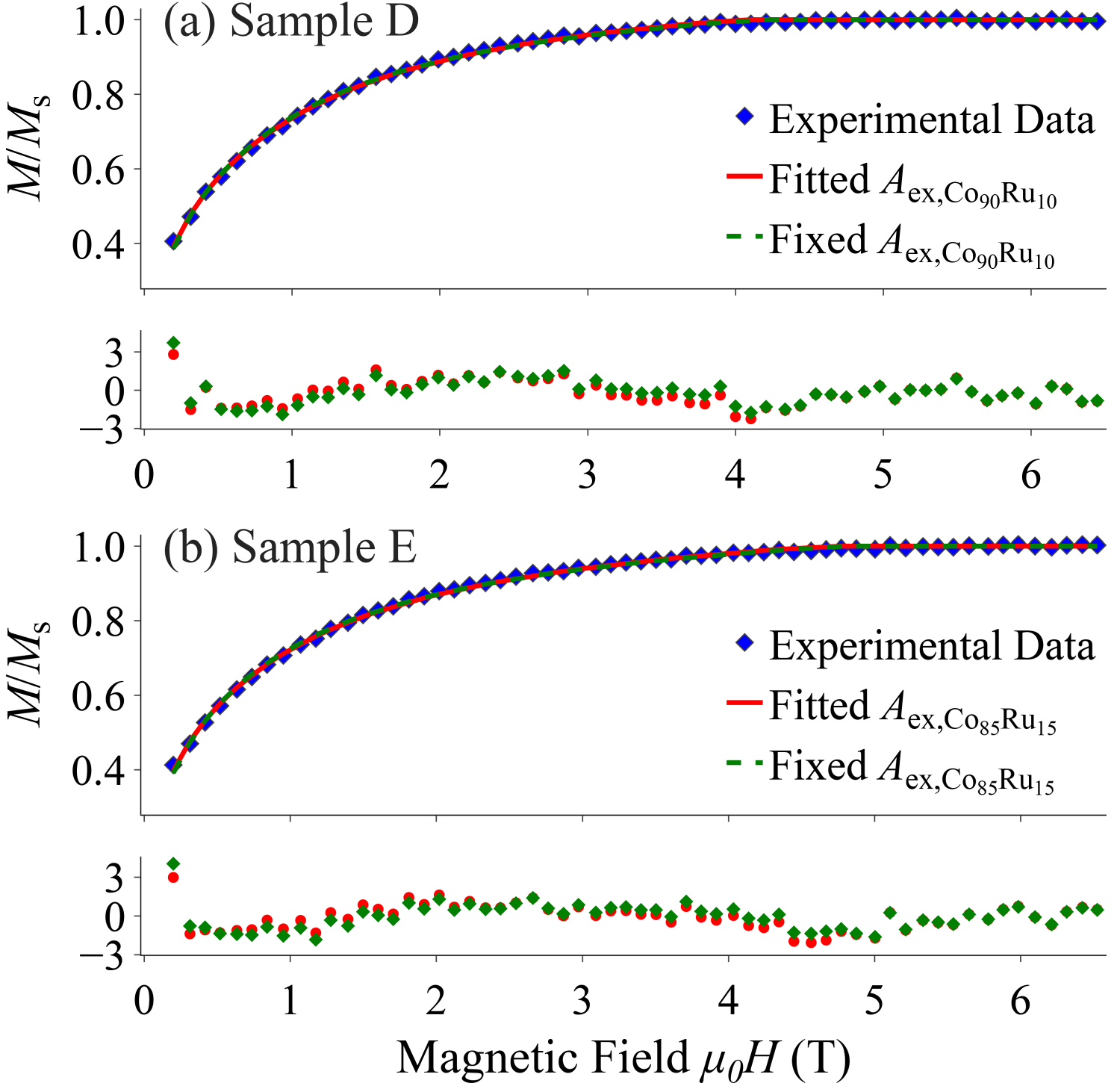}
    \caption{Fits of $M(H)$ curves for (a) sample D and (b) sample E using the discrete energy model. In both fits, $J_1$ and $J_2$ are the fitting parameters, while $\A{\mathrm{Co}_x \mathrm{Ru}_{1-x}}$ is either a constant or treated as a fitting parameter. $A_{\mathrm{ex, Co}}$ is constrained to \SI{27.5}{\pico\joule\per\meter}. The transition stiffness in both fits is set to that of the Co$_x$Ru$_{1-x}$ layer. Below each fit, standardized residuals are shown for both the constrained and free $\A{\mathrm{Co}_x \mathrm{Ru}_{1-x}}$ cases.}
    \label{fig: D vs E}
\end{figure}

As discussed in \cref{sec: Theory}, it is necessary to have relatively large interlayer coupling between in FM$_1$/SL/FM$_1$ to measure $A_\mathrm{ex}$ by fitting to $M(H)$ data. While Ru spacer layers are known to be able to produce large interlayer coupling between Co, CoFe, and CoFeB layers, they fail to do so for many other alloys and materials, e.g. those with low $M_s$. In these cases, one can introduce atomically-thin coupling layers between the FM layers and the SL which then act to induce a large spin spiral when an external magnetic field is applied. This approach is reproduced in samples D and E, where a \SI{0.8}{\nano\meter} Co layer is inserted between Co$_x$Ru$_{1-x}$ and the SL. We provide two different fits of these structures using the discrete energy model: with the stiffness of the Co$_x$Ru$_{1-x}$ layer fixed to the value obtained in \cref{tbl: stiffnesses} and with this stiffness as a free fitting parameter. In both cases, the stiffness of the interfacial Co layer is fixed to the fitted value of $\SI{27.5}{\pico\joule\per\meter}$. These fits are shown in \Cref{fig: D vs E}. 

When left as fitting parameter, we obtain best-fit values of $\A{\mathrm{\coninety}} = \SI{15.4 \pm 0.5}{\pico\joule\per\meter}$  (sample D) and $\A{\mathrm{\coeighty}} = \SI{8.4 \pm 0.2}{\pico\joule\per\meter}$ (sample E). These can be compared to the values obtained by fitting $\A{\mathrm{Co}_x \mathrm{Ru}_{1-x}}$ in structures without the coupling layer: $\A{\mathrm{\coninety}} = \SI{13.0 \pm 0.4}{\pico\joule\per\meter}$  (sample B) and $\A{\mathrm{\coeighty}} = \SI{7.2 \pm 0.4}{\pico\joule\per\meter}$  (sample C). Both values of $\A{\mathrm{Co}_x \mathrm{Ru}_{1-x}}$ increase slightly when fitted, but this may be affected by the choice of transition stiffness being set to the stiffness of the outer layer, Co$_x$Ru$_{1-x}$. If, instead, the transition stiffness is taken to be equal to $\A{\mathrm{Co}}$, we see best-fit values of $\A{\mathrm{\coninety}} = \SI{14.1 \pm 0.5}{\pico\joule\per\meter}$ (sample D) and $\A{\mathrm{\coninety}} = \SI{7.3 \pm 0.2}{\pico\joule\per\meter}$ (sample E), which are in good agreement with the values for samples B and C in \cref{tbl: stiffnesses}. This suggests that when using an interfacial layer to measure the stiffness of low-coupling layers, one should consider using each transition stiffness to establish bounds on the proper value of $A_\mathrm{ex}$. 

In all cases, the best-fit values of $J_1$ and $J_2$ for samples D and E with $\A{\mathrm{Co}_x \mathrm{Ru}_{1-x}}$ as a free parameter are within 2\% of those reported in \cref{tbl: fits}, where it is fixed. 

%%%%%%%%%%%%%%%%%%%%%%%%%%%%%%%%%%%%%%%%%%%%%%%%%%%%%%%%%%%%%%
% DISCUSSION
%%%%%%%%%%%%%%%%%%%%%%%%%%%%%%%%%%%%%%%%%%%%%%%%%%%%%%%%%%%%%%

\section{\label{Sec: Discussion}Conclusion}

This paper discusses three distinct models to describe the magnetization reversal in magnetic multilayers with antiferromagnetic or noncollinear coupling. These models assume either a discrete or continuous sub-structure in the ferromagnetic layer, and either try to minimize the total magnetic energy or fix the total magnetic torque at zero. The theory of each model was presented in detail, and it was shown that the discrete torque model can be manipulated to mathematically reduce to the continuous torque model. While the energy and continuous torque models were introduced in previous works \cite{girt_method_2011, omelchenko_continuous_2022}, the methods employed to generalize these models to asymmetric and multi-ferromagnetic structures are novel, along with the entire formulation of the discrete torque model. 

We validated these three models by fitting them to the experimentally measured $M(H)$ curves of a range of symmetric single-ferromagnetic, symmetric multi-ferromagnetic, and asymmetric multi-ferromagnetic structures. These results show that computational performance varies widely across the proposed models, and run-times especially scale poorly with structure complexity for the continuous model. We showed that, while best-fit values of the interlayer exchange coupling constants are largely the same across the three models, there is a statistically significant improvement in agreement between the discrete models in comparison with the continuous for the multi-ferromagnetic and asymmetric structures. As well, we see that quality of fit, as measured by $\chi^2_\nu$, is consistently better for the discrete models for samples D--G. Our detailed exploration of the spin structure models allowed us to gain insight and offer an explanation for the discrepancy between the newly reported values and literature values of the exchange stiffness of Co layers: the interfacial layers of reduced stiffness strongly affect the overall fitted value of $A_\mathrm{ex}$, but not the best-fit $J_1$ and $J_2$. We show that the choice of SL material, e.g. using RuFe instead of Ru, can minimize this effect by lowering the concentration of nonmagnetic material in SL. Finally, we demonstrate how one can introduce interfacial coupling layers to fit samples with low interlayer coupling. 

All three proposed models produce identical $\chi^2_\nu$ for the fits of the symmetric samples, A--C. As well, best-fit values of $J_1$ and $J_2$ for these samples are equal to within the standard error, while $A_\mathrm{ex}$ differs slightly only for sample A. Thus, our only metric to performatively distinguish the models for symmetric single-ferromagnetic-layered structures is computational run-time, and the continuous torque model reaches a minimized $\chi^2_\nu$ about an order of magnitude faster than the discrete models. By applying the same assumptions of the symmetric continuous model to the discrete torque model, this performance gap could be closed for symmetric structures. For multi-ferromagnetic samples, the discrete models are more appropriate. For samples D--G, $\chi^2_\nu$ is larger for the continuous model and best-fit values of $J_1$ and $J_2$ differ from the discrete models by as much as 22\% and 6\%, respectively. Additionally, run-times for the continuous torque model are prohibitive for complex structures, taking an order of magnitude longer to fit for sample G. For arbitrary asymmetric samples, we found that the discrete energy model is most appropriate for $N + \mathcal{N} < 42$, as it has marginally lower run-times than the discrete torque, but produces similar fit statistics. For asymmetric structures with $N + \mathcal{N} > 42$, we showed that the discrete torque model exhibits improved run-time scaling. 

One can directly measure $A_\mathrm{ex}$ through Brillouin light and neutron scattering experiments, but these methods may not be accessible to many laboratories. Instead, one can also indirectly measure $A_\mathrm{ex}$ by fitting to the $M(H)$ of strongly antiferromagnetically or noncollinearly coupled multilayers, where a spin spiral across the SL is generated by the interlayer exchange coupling when an external magnetic field is applied, which is necessary for VSM measurements. However, strong antiferromagnetic coupling can still be produced in weakly-coupled structures by inserting a thin Co layer at the interfaces of the SL, which induces the necessary spin spiral. This approach enables the design of structures for measuring the stiffness of magnetic layers with small or no antiferromagnetic or noncollinear coupling across spacer layers like Ru. 

The computational performance of the discrete torque model is limited by the issues discussed in \cref{app: T solving}. If these were addressed, the discrete torque model may outperform the discrete energy for any $N + \mathcal{N}$. A combination of approaches might be employed instead of a pure torque model, wherein the energy model is used to facilitate a gradient descent approach to efficiently explore the solution space, rather than relying on iterative rejection of non-propagating solutions. 

\section{\label{sec: data availability}Data and model availability}

We provide a public-access website accessible via the ``Tools'' section at \href{https:/www.sfu.ca/pnmd.html}{\url{sfu.ca/pnmd}} where users can apply our discrete energy model to $M(H)$ data. The data used to generate \cref{fig: calibration fits,fig: M(H) fits} in this article are available on this site. This tool is provided as-is and is available for external users to upload and fit their own data for arbitrary antiferromagnetic or noncollinearly-coupled multilayers. 

%%%%%%%%%%%%%%%%%%%%%%%%%%%%%%%%%%%%%%%%%%%%%%%%%%%%%%%%%%%%%%
% ACKNOWLEDGEMENTS
%%%%%%%%%%%%%%%%%%%%%%%%%%%%%%%%%%%%%%%%%%%%%%%%%%%%%%%%%%%%%%

\bigskip

\begin{acknowledgments}

We acknowledge the support of the Natural Sciences and Engineering Research Council of Canada (NSERC), [funding reference number RGPIN-2019-07203]. 

E.W., A.T., and G.L.-L. contributed equally to this work.

\end{acknowledgments}

\bibliography{references}

%%%%%%%%%%%%%%%%%%%%%%%%%%%%%%%%%%%%%%%%%%%%%%%%%%%%%%%%%%%%%%
% APPENDIX
%%%%%%%%%%%%%%%%%%%%%%%%%%%%%%%%%%%%%%%%%%%%%%%%%%%%%%%%%%%%%%

\appendix
\section{\label{app: torque}Exchange Torque Derivation}
% Appendix A: Exchange Torque Derivation

If we consider two adjacent spins labeled $\spin{1}$ and $\spin{2}$, the exchange energy may be expressed as
\begin{equation}
    E_{\mathrm{ex}} = -J\spin{1}\cdot \spin{2},
    \label{eq: exchange energy}
\end{equation}
where the relation between $\spin{1,2}$ and the magnetic moments, $\vec{\mu}_{1,2}$, is as follows:
\begin{equation}
    \vec{\mu}_{1,2} = g\frac{\mu_B}{\hbar} \spin{1,2},
    \label{eq: g-factor}
\end{equation}
where $g$ is the electronic g-factor and $\mu_B$ is the Bohr magneton ($\mu_B = e\hbar/2m_e$,
where $e$ and $m_{e}$ are the charge and mass of an electron, respectively).
For spins arranged in a simple cubic structure, the volumetric magnetization density, $M_{s1,2}$, is given by:
\begin{equation}
    \Ms{1,2} = \dfrac{|\vec{\mu}_{1,2}|}{a^3} = |\M{1,2}|,
    \label{eq: Ms}
\end{equation}
where $a$ is the atomic layer spacing. To calculate the torque on $M_1$, we need to calculate the intermolecular field generated by $M_2$ on $M_1$, $\Hto$. The relation between this field and the exchange energy of a magnetic moment $\spin{1}$ is given by
\begin{align}
    -\vec{\mu}_{1} \cdot \mu_0\Hto &= E_{ex}, \\
    \vec{\mu}_{1} \cdot \mu_0\Hto &= J\spin{1}\cdot \spin{2}. \\
\intertext{Substituting $\spin{1}$ from \cref{eq: g-factor} gives}
   \vec{\mu}_{1} \cdot \mu_0\Hto &= J \dfrac{\hbar}{g\mu_B} \vec{\mu}_{1} \cdot \spin{2},\\
\intertext{which implies}
    \mu_0\Hto &= J \dfrac{\hbar}{g\mu_B} \spin{2}.
    \label{eq: mu0 H}
\end{align}
Next, combining \cref{eq: g-factor,eq: Ms}, gives
\begin{equation}
   \dfrac{\hbar}{g\mu_B} = \dfrac{|\spin{1}|}{a^3\Ms{1}}.
\end{equation}
Substituting this into \cref{eq: mu0 H}) results in
\begin{equation}
    \mu_0\Hto = J \dfrac{|\spin{1}|}{a^3\Ms{1}}\spin{2}.
    \label{eq: second to last step}
\end{equation}
Finally, we note that $\spin{2}$ and $\M{2}$ are in the same direction, and therefore we may write
\begin{equation}
    \spin{2} = |\spin{2}|\dfrac{\M{2}}{\Ms{2}},
\end{equation}
which, after substitution into (\cref{eq: second to last step}), gives the final expression for $\Hto$ as
\begin{equation}
    \mu_0\Hto = J\dfrac{|\spin{1}||\spin{2}|}{a^3}\dfrac{\M{2}}{\Ms{1}\Ms{2}}.
\end{equation}
Therefore, the torque on $\M{1}$ from $\M{2}$ can be calculated as
\begin{equation}
\begin{aligned}
    \Tto &= a\M{1} \times \mu_0\Hto, \\
    &=  2\dfrac{\A{1(2)}}{a}\dfrac{\M{1} \times \M{2}}{\Ms{1}\Ms{2}},
\end{aligned}
\end{equation}
where $\A{1(2)}$ is the exchange stiffness between two magnetic atomic layers 1 and 2 and is defined as 
\begin{equation}
    \A{1(2)} = J\dfrac{|\spin{1}||\spin{2}|}{a}.
\end{equation}

\section{\label{app: transition stiffness}Transition Stiffness}
% \section{Transition Stiffness}

To use the proposed models to fit the measured $M(H)$ curves of structures with multiple layers of \textit{distinct} ferromagnetic materials on either side of the spacer layer, one must know the stiffness between each pair of ferromagnetic layers that are in contact. This ``transition stiffness'' cannot be determined experimentally, however, it can be fit as a free parameter in the discrete models. This appendix will demonstrate the dependence of the $M(H)$ curve on the choice of value of transition stiffness for different numbers of atomic layers.

For simplicity, we consider a symmetric FM$_1$/\allowbreak FM$_2$/\allowbreak SL/\allowbreak FM$_2$/FM$_1$ multilayer structure where FM$_1$ and FM$_2$ on both sides of the spacer layer have the same thickness. The FM layers have thicknesses of 0.4, 1, and \SI{2}{\nano\meter}, with the thickness of a single magnetic atomic layer kept constant at \SI{0.2}{\nano\meter}. Thus, each FM layer has either two, five, or 10 magnetic atomic planes. The exchange stiffness of FM$_1$ is chosen to be five times smaller than that of FM$_2$, $\A{1}=\SI{30}{\pico\joule\per\meter}$, $\A{2}=\SI{6}{\pico\joule\per\meter}$, to magnify the effect induced by the difference between the exchange stiffnesses of the magnetic layers. The values of the interlayer exchange coupling constants are chosen to be similar to those obtained from experimentally studied samples with Co interface layers, $J_1=\SI{3.0}{\milli\joule\per\meter\squared}$ and $J_2=\SI{1.5}{\milli\joule\per\meter\squared}$.

\begin{figure}[htbp]
    \includegraphics[width=1.0\columnwidth]{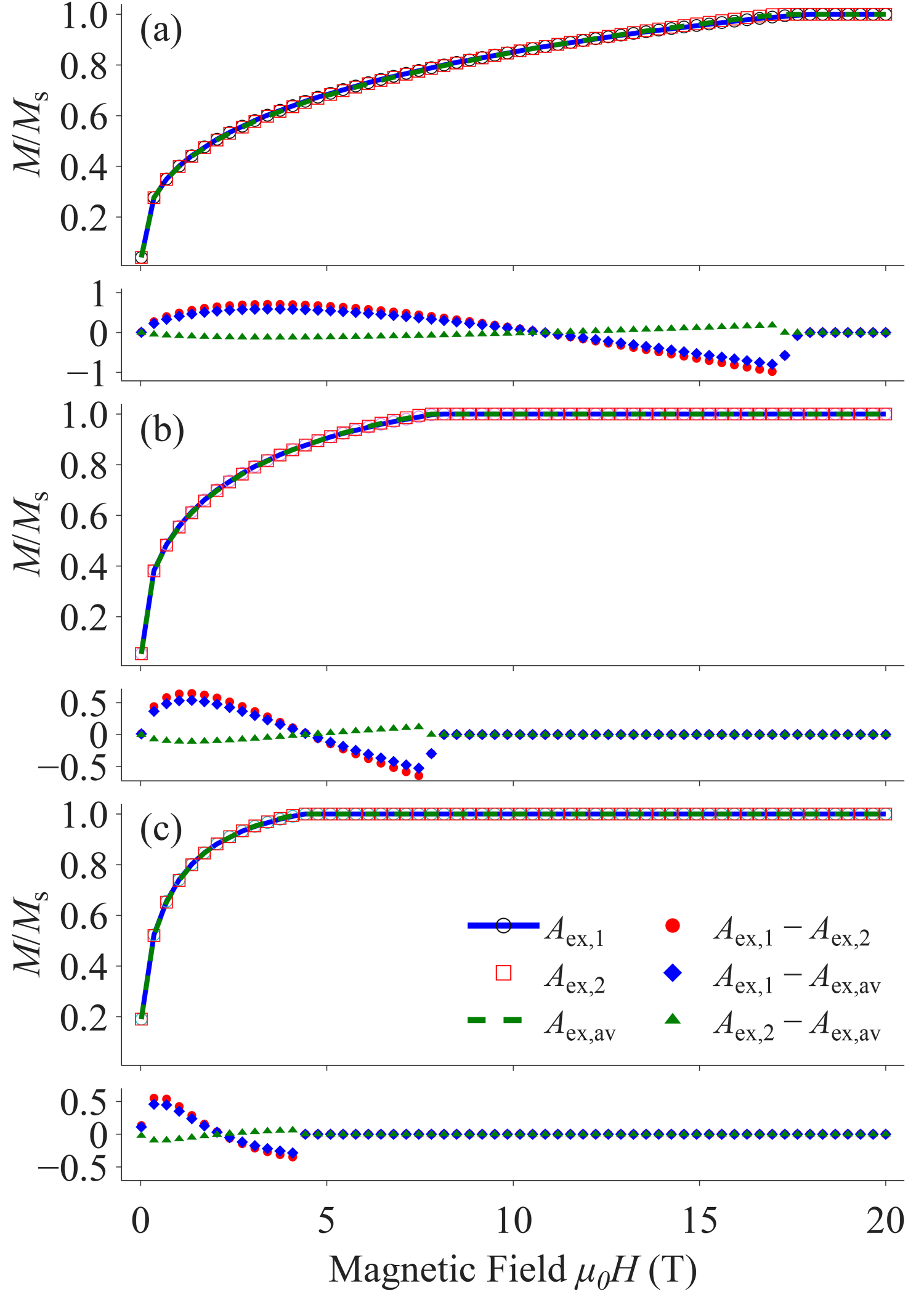}
    \caption{$M(H)$ curves simulated using different transition stiffnesses $\A{1}$, $\A{2}$, and $\A{\mathrm{av}}$, where $\A{\mathrm{av}} = (\A{1} +\A{2})/2$. Each magnetic layer has a) two, b) five, or c) 10 magnetic atomic layers. The percent difference shown in the sub-figure of each plot is calculated by subtracting two given $M(H)$ curves (as shown in the legend), normalizing with the $M(H)$ values obtained using a transition stiffness of $\A{\mathrm{av}}$.}
    \label{fig: transitions}
\end{figure}

As shown in each sub-figure of \Cref{fig: transitions}, there is at most a 1\% difference between the simulated $M(H)$ curves obtained using a transition stiffness of $\A{1}$, $\A{2}$, and the average, $\A{\mathrm{av}}=(\A{1} + \A{2})/2$. With an increasing number of magnetic atomic planes, the transition stiffness has a smaller impact on the behavior of $M(H)$. It is also shown that there is a smaller difference between $M(H)$ curves obtained by using $\A{2}$ and $\A{\mathrm{av}}$ than $\A{1}$ and $\A{\mathrm{av}}$ transition stiffnesses. 

\begin{figure}[htbp]
    \includegraphics[width=1.0\columnwidth]{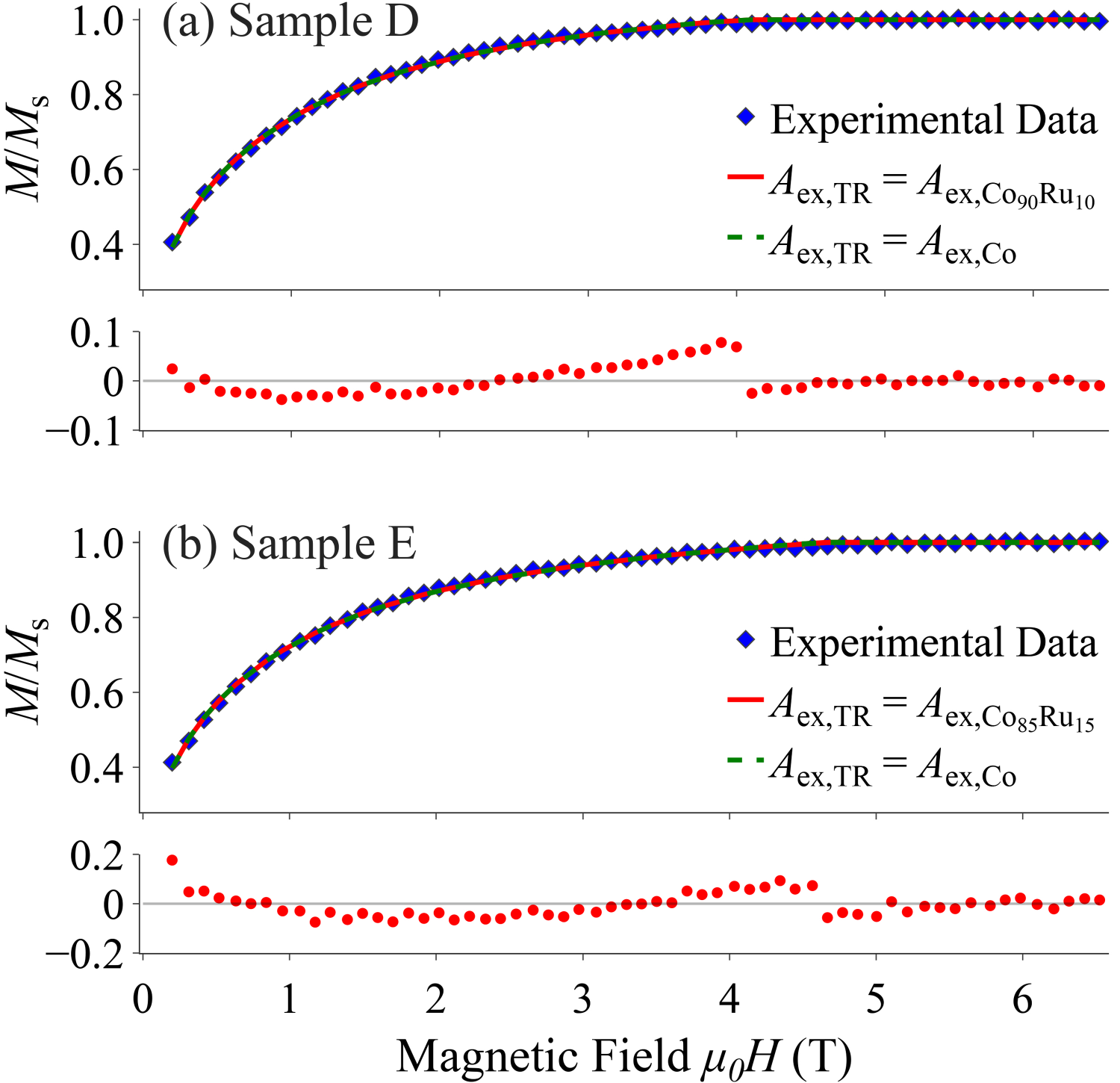}
    \caption{Fits of $M(H)$ curves of (a) sample D and (b) sample E using the discrete energy model. In these fits, $J_1$, $J_2$, and $\A{\mathrm{Co_x Ru_{1-x}}}$ are the fitting parameters, while $\A{\mathrm{Co}}$ was constrained to \SI{27.5}{\pico\joule\per\meter}. For each sample, the fitting was performed with the transition stiffness, $\A{\mathrm{TR}}$, set either to the stiffness of the Co layer ($\A{\mathrm{TR}} = \A{\mathrm{Co}} = \SI{27.5}{\pico\joule\per\meter}$) or to that of the Co$_x$Ru$_{1-x}$ layer ($\A{\mathrm{TR}} = \A{\mathrm{Co_x Ru_{1-x}}}$). Each figure is accompanied by a sub-figure illustrating the difference in standardized residuals for fits with $\A{\mathrm{TR}} = \A{\mathrm{Co}}$ and $\A{\mathrm{TR}} = \A{\mathrm{Co_x Ru_{1-x}}}$ transition stiffnesses.}
    \label{fig: transitions2}
\end{figure}

To demonstrate how the transition stiffness affects the best-fit values of $J_1$, $J_2$, and $A_{\mathrm{ex}}$, we also fit samples D and E, see \cref{tbl:samples}, which are symmetric structures (Co$_x$Ru$_{1-x}$/Co/SL/Co/Co$_x$Ru$_{1-x}$) with a large discrepancy in exchange stiffness between the Co and Co$_x$Ru$_{1-x}$ layers. In these fits, $J_1$, $J_2$, and $\A{\mathrm{CoRu}}$ are the fitting parameters, while $\A{\mathrm{Co}}$ is constrained to \SI{27.5}{\pico\joule\per\meter} (see \cref{tbl: stiffnesses}). The transition stiffness, $\A{\mathrm{TR}}$, is either set to the exchange stiffness of Co ($\A{\mathrm{Co}} = \SI{27.5}{\pico\joule\per\meter}$) or to that of the Co$_x$Ru$_{1-x}$ layers (a fitted parameter). These fits and the corresponding experimental data are illustrated in \Cref{fig: transitions2}, the best-fit values of $A_\mathrm{ex}$, $J_1$, and $J_2$ are given in \cref{tbl: transition stiff fits}. 
\begin{table}[htbp]
    \begin{ruledtabular}
        \begin{tabular}{c|ccc}
            \textbf{Sample}  & \begin{tabular}[c]{@{}c@{}}{\boldmath{${A_\mathrm{ex}}$}} \\[-1ex] 
            {$\mathbf{(\SI{}{\pico\joule\per\meter})}$}\end{tabular} & \begin{tabular}[c]{@{}c@{}}{\boldmath{$J_1$}}\\[-1ex] {$\mathbf{(\SI{}{\milli\joule \per\meter\squared})}$}\end{tabular} & \begin{tabular}[c]{@{}c@{}}{\boldmath{$J_2$}}\\[-1ex] {$\mathbf{(\SI{}{\milli\joule\per\meter\squared})}$}\end{tabular} \\ \hline
            D ({$\A{\mathrm{TR}} = \A{\mathrm{Co}}$})   & 14.1(5) & 3.333(7) & 1.466(10) \\
            D ({$\A{\mathrm{TR}} = \A{\mathrm{\coninety}}$}) & 15.4(5) & 3.333(7) & 1.466(10) \\ \hline
            E ({$\A{\mathrm{TR}} = \A{\mathrm{Co}}$})   & 7.3(2)  & 2.873(7) & 1.332(10) \\
            E ({$\A{\mathrm{TR}} = \A{\mathrm{\coeighty}}$}) & 8.4(2)  & 2.873(7) & 1.331(10)
        \end{tabular}
    \end{ruledtabular}
    \caption{Best-fit values of exchange stiffness and the interlayer coupling constants for samples D and E, either using a transition stiffness, $\A{\mathrm{TR}}$, equal to the stiffness of Co ($\A{\mathrm{Co}}$ = \SI{27.5}{\pico\joule\per\meter}) or equal to that of the Co$_x$Ru$_{1-x}$ layer, which is a free fitting parameter.}
    \label{tbl: transition stiff fits}
\end{table}

It is evident from the best-fit values that choosing the transitional stiffness to be either that of Co or Co$_x$Ru$_{1-x}$ does not affect the resulting coupling constants. In both cases, $J_1 = \SI{3.333 \pm 0.007}{\milli\joule\per\square\meter}$ and $J_2 = \SI{1.466 \pm 0.01}{\milli\joule\per\square\meter}$. However, the choice of transition stiffness has a slight impact on the fitted value of $\A{\mathrm{CoRu}}$, and the resulting values of $A_\mathrm{ex}$ do not agree to within the standard error. Thus, when using FM$_2$ as an interfacial coupling layer to measure the stiffness of a low-coupling FM$_1$ in FM$_1$/FM$_2$/SL/FM$_2$/FM$_1$, one could perform two fits: one with the transition stiffness equal to the exchange stiffness of FM$_2$ and another with it set to the stiffness of FM$_1$, now as a free fitting parameter. This approach would allow for an accurate determination of the uncertainty in the measurement of the stiffness of FM$_1$.

Thus, as $A_\mathrm{ex}$ is not treated as a free parameter in the fitting of the multi-ferromagnetic and asymmetric structures in the rest of this work, we elect to use a transition stiffness equal to the bulk exchange stiffness of the outer layer, FM$_1$, as it has been shown to have no effect on the best-fit values of $J_1$ and $J_2$.

\medskip

\section{\label{app: T solving}Solving the Discrete Torque Model}
% \section{Solving the Discrete Torque Model}

The torque model proposes \cref{eq: T bulk,eq: T i=1,eq: T i=t,eq: T i=N,eq: T i=N+1}, which must all be satisfied to determine the distribution of angles, $\theta_i$, within the structure. As there is no clear path toward an analytical solution, one must apply numerical methods to solve these equations. This is not as straightforward as for the discrete energy model, where one can simply run a minimization routine on the energy function to determine $\theta_i$. Here, instead, we apply the ``shooting method'' for solving this boundary value problem.

A random starting angle $\theta_1$ is chosen, from which a $\theta_2$ can be calculated using a rearranged form of \cref{eq: T i=1}. With this arbitrary pair of $\theta_1$ and $\theta_2$, one can calculate all the angles up to $\theta_{N}$ using \cref{eq: T bulk}. From here, \cref{eq: T i=N+1} may be solved numerically for the values of $\theta_{N+1}$. There may be multiple solutions or no solution to this equation, depending on the choice of starting angle $\theta_1$. Next, combining \cref{eq: T i=N} and \cref{eq: T i=N+1}, we arrive at a new equation for $\theta_{N+2}$ in terms of known values:
\begin{equation}
\begin{aligned}
    &\sin(\theta_{N+1} - \theta_{N+2}) = \\
    & \dfrac{a_{(N+1)(N+2)}\A{N(N-1)}}{a_{(N)(N-1)}\A{(N+1)(N+2)}}\sin(\theta_{N} - \theta_{N-1}) \\ 
    + &\dfrac{a_{(N+1)(N+2)}}{2\A{(N+1)(N+2)}}\left[a_{(N)(N-1)}\Ms{N}\sin(\theta_N) \right. \\ 
    - &\left.a_{(N+1)(N+2)}\Ms{N+1}\sin(\theta_{N+1})\right].
\end{aligned}
\end{equation}
With a solution to both $\theta_{N+1}$ and $\theta_{N+2}$, the remaining angles can be determined using \cref{eq: T bulk}. At this point, all values of $\theta_{i}$ have been found. Finally, these values of $\theta_{i}$ are verified using \cref{eq: T i=t}. If this condition is not satisfied, the process is restarted with another starting angle. Thus, the problem is effectively reduced to one dimension. 

To handle the multiple solutions when solving for $\theta_{N+1}$, one bounds the possible solutions to the range $[0, \pi]$, in which case there should be, at most, two solutions. Both solutions must be propagated to the opposite interface ($N + \mathcal{N}$), and the one that remains within the range [0, 2$\pi$] is taken as physical. For each physical solution for a given $\theta_1$, one then applies a specialized root finding algorithm to find a solution to \cref{eq: T i=N+1} at the interface and, if a null result is returned, no information is gained. In these cases, the solving algorithm is required to restart with another guess for $\theta_1$. As large ranges of values of starting angles may return null results, this significantly reduces the potential performance of the discrete torque model. Despite having potentially orders of magnitude more free parameters, the energy model thus vastly outperforms the discrete torque model when fitting $M(H)$ with arbitrary numbers of layers. Also, note that the symmetric models (\citep{omelchenko_continuous_2022, girt_method_2011}) do not present this issue, as one can simply assume a symmetric $\theta_i$ distribution on either side of the SL. This explains the great decrease in computation times observed by  \citet{omelchenko_continuous_2022}.

\end{document}